%% file: main.tex
\documentclass[aps,prb,twocolumn,groupedaddress,superscriptaddress,amsfonts,amssymb,amsmath,citeautoscript,longbibliography,a4paper]{revtex4-2}
\usepackage{layouts}
\input{setup}

\graphicspath{{figs/}}

\setboolean{togglecomments}{true}
\setboolean{toggletodos}{true}

\setboolean{togglechanges}{false}

\begin{document}
\title{Symbolic Learning of Topological Bands in Photonic Crystals}


\newcommand{\mitaffil}{Department of Physics, Massachusetts Institute of Technology, Cambridge, MA 02139, USA}
\newcommand{\yaleaffil}{Department of Applied Physics, Yale University, New Haven, Connecticut 06520, USA}
\newcommand{\iaifiaffil}{The NSF Institute for Artificial Intelligence and Fundamental Interactions}
\newcommand{\rleaffil}{Research Laboratory of Electronics, Massachusetts Institute of Technology, Cambridge, Massachusetts 02139, USA}
\newcommand{\dtuaffil}{Department of Electrical and Photonics Engineering, Technical University of Denmark, 2800 Kgs.\ Lyngby, Denmark\\
$^\bigstar$ denotes equal contribution}

\author{Ali~Ghorashi$^\bigstar$}
\email{aligho@mit.edu}
\affiliation{\mitaffil}
\affiliation{\yaleaffil}

\author{Sachin~Vaidya$^\bigstar$}
\email{svaidya1@mit.edu}
\affiliation{\mitaffil}
\affiliation{\rleaffil}
\affiliation{\iaifiaffil}

\author{Ziming Liu}
\affiliation{\mitaffil}
\affiliation{\iaifiaffil}

\author{Charlotte Loh}
\affiliation{\mitaffil}

\author{Thomas Christensen}
\affiliation{\dtuaffil}

\author{Max Tegmark}
\affiliation{\mitaffil}
\affiliation{\iaifiaffil}

\author{Marin~Solja\v{c}i\'c}
\affiliation{\mitaffil}
\affiliation{\rleaffil}
\affiliation{\iaifiaffil}


\begin{abstract}  
Topological photonic crystals (PhCs) that support disorder-resistant modes, protected degeneracies, and robust transport have recently been explored for applications in waveguiding, optical isolation, light trapping, and lasing. However, designing PhCs with prescribed topological properties remains challenging because of the highly nonlinear mapping from the continuous real-space design of PhCs to the discrete output space of band topology. Here, we introduce a machine learning approach to address this problem, employing Kolmogorov--Arnold networks (KANs) to predict and inversely design the band symmetries of two-dimensional PhCs with two-fold rotational ($C_2$) symmetry. We show that a single-hidden-layer KAN, trained on a dataset of $C_2$-symmetric unit cells, achieves high accuracy in classifying the topological classes of the lowest lying bands. We use the symbolic regression capabilities of KANs to extract algebraic formulas that express the topological classes directly in terms of the Fourier components of the dielectric function. These formulas not only retain the full predictive power of the network but also provide novel insights and enable deterministic inverse design. Using this approach, we generate photonic crystals with target topological bands, achieving high accuracy even for high-contrast, experimentally realizable structures beyond the training domain.
\end{abstract}
\maketitle


Topological photonic crystals (PhCs) have rapidly emerged as highly promising platforms for achieving robust, disorder-resistant light confinement and transport~\cite{wang2009observation, rechtsman2013photonic, skirlo2015experimental, schulz2021topological}. They exhibit unique features including unidirectional edge states, defect-localized modes, and topological degeneracies that are resilient against fabrication imperfections and perturbations. Such properties make topological PhCs potentially useful candidates for next-generation devices in waveguiding, lasing, optical isolation, and integrated photonics~\cite{lu2014topological, ozawa2019topological, jalali2023topological, smirnova2020nonlinear, ota2020active, schulz2021topological}. Following early realizations of fundamental topological phenomena, such as Chern and Weyl phases~\cite{lu2013weyl, haldane2008possible, wang2009observation, rechtsman2013photonic, vaidya2020observation}, the field has expanded to include higher-order~\cite{he2020quadrupole, xie2018second}, non-Hermitian~\cite{parto2020non, pontula2024non}, quantum~\cite{jalali2023topological, sloan2025noise}, and nonlinear~\cite{smirnova2020nonlinear, jurgensen2021quantized, mukherjee2021observation} topological physics.

Despite significant progress in theoretical and survey-based explorations of the topological design space~\cite{ghorashi2024prevalence, vaidya2023topological, bradlyn2017topological, po2018fragile, song2020fragile, xie2018second, he2020quadrupole, davoodi2025active}, the inverse design of PhCs with prescribed topological bands remains a fundamental challenge~\cite{kim2023automated, christiansen2021inverse}. This difficulty arises primarily from the inherent complexity of the mapping from continuous real-space geometry of the PhCs (the dielectric function) to the discrete, integer-valued output space of band topology. Effectively, there is an absence of general, physically intuitive principles that universally link the geometric parameters of a PhC to the topological classification of its bands. Conventional inverse design methods often rely on heuristic parameter sweeps, expensive random searches, or black-box optimization techniques, which offer limited physical insight and can struggle with the high-dimensional nature of the design space.

In recent years, machine learning (ML) has shown great potential in addressing this design complexity. Convolutional Neural Networks (CNNs), motivated by success in image recognition~\cite{rawat2017deep, guo2017simple}, have been effectively applied to predict the band structures of two-dimensional PhCs~\cite{christensen2020predictive} and topological invariants, such as winding numbers in one-dimensional models~\cite{zhang2018machine}. Furthermore, Generative Adversarial Networks (GANs) and other deep learning architectures have been employed to optimize bandgaps~\cite{christensen2020predictive} and tailor edge state dispersions~\cite{pilozzi2018machine}. While these models achieve high predictive accuracies, they are predominantly black-box in nature. Although post-hoc analysis of a trained network's weights can sometimes reveal implicit physical laws~\cite{zhang2018machine}, this process is often non-generalizable, limiting the direct extraction of analytical design rules. The lack of explicit formulas is particularly problematic in inverse design, where constraining the solution space is critical to avoid the pervasive many-to-one problem~\cite{deng2024inverse, chen2022inverse, jiang2021deep, gupta2023tandem, jiang2025machine}.

Here, we pursue two complementary objectives to enable the design of topological photonic bands: (1) to develop a machine learning model that yields interpretable phenomenological laws mapping PhC geometries directly to the topological classification of their bands, and (2) to leverage this model for the inverse design of PhCs with deterministic topological bands. We focus on the simplest nontrivial plane group, \emph{p}2, in which the two-fold rotation ($C_2$) eigenvalues at high-symmetry points help diagnose the topological classification of the bands. In this setting, the PhCs can host a plethora of topological effects including polarization-induced edge and corner states, Chern bands with chiral edge states, and Dirac points.

To realize these objectives, we employ the recently introduced Kolmogorov-Arnold networks (KANs)~\cite{liu2024kan} architecture trained on a dataset of PhCs spanning continuous variations in real-space geometry. We augment this dataset using insights from topological quantum chemistry~\cite{kruthoff2017topological, bradlyn2017topological, po2017symmetry} to increase the data available for training. The trained model achieves high predictive accuracy ($>99\%$) and identifies that only the smallest Fourier components of the dielectric function govern the band topology of the lowest bands, consistent with expectations from degenerate perturbation theory. The model also compresses the mapping from real-space geometry, expressed as Fourier coefficients, and band symmetry into analytic expressions through symbolic regression. Finally, we demonstrate an inverse design strategy that target specific topological classes using these expressions. We use these explicit formulas for the deterministic inverse design of experimentally realizable, two-tone, high-contrast PhCs, belonging to specific topological classes, which notably lie beyond the training domain.

\begin{figure}
    \centerline{%
    \includegraphics[scale=1]{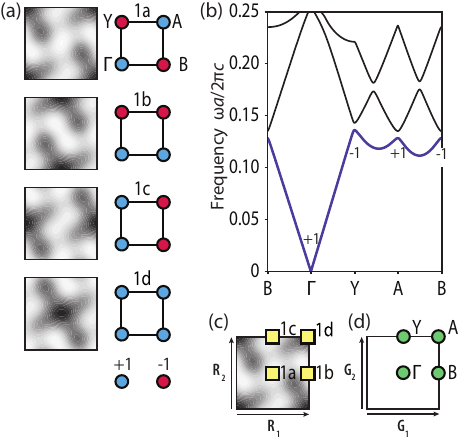}}
    \caption{%
        (a)~Dielectric profile of a typical photonic crystal in our dataset, depicted in the basis of lattice vectors, $\mathbf{R}_1, \mathbf{R}_2$. Our dataset is augmented four-fold by considering shifts of the unit cell center to four $C_2$ symmetric Wyckoff positions: 1a, 1b, 1c and 1d (see subfigure (c)), which change  the symmetry eigenvalues according to \cref{equation; data augmentation}. In the left column of (a), we show the same photonic crystal at these four choices of centering and in the right column we show the corresponding symmetry eigenvalues at the four high symmetry points in the Brillouin zone, with blue and red circles denoting $+1$ and $-1$ $C_2$ eigenvalues, respectively.  
        (b)~TM-polarized band dispersion of the PhC in (a). The first band is highlighted in blue. We indicate the $C_2$ symmetry eigenvalues of the first band, corresponding to centering at the 1a Wyckoff position. (c) Position of the Wyckoff positions (yellow squares) in the unit cell overlaid on the PhC centered at 1a. (d) High symmetry points in the Brillouin zone (green circles) shown in the basis of primitive reciprocal lattice vectors, $\mathbf{G}_1, \mathbf{G}_2$. 
        }
    \label{figure: 1}
\end{figure}

\paragraph{Topological phases in $C_2$-symmetric PhCs}
Photonic bands in a $C_{2}$-symmetric crystal admit a symmetry-based classification that follows from the eigenvalues of the two-fold rotation operator ($C_2$) at high-symmetry momenta. At the high-symmetry points $\Gamma$, $B$, $Y$, and $A$ in the Brillouin zone, each transverse-magnetic (TM) or transverse-electric (TE) mode transforms according to one of the two one-dimensional irreducible representations of $C_{2}$, with eigenvalues $+1$ or $-1$ (\cref{figure: 1}). For the lowest band, the eigenvalue at $\Gamma$ is fixed to +1 by continuity~\cite{christensen2022location}, which leaves eight possible combinations of eigenvalues. These combinations may be separated into two categories according to the parity of the number of $-1$ eigenvalues at the four high-symmetry points. Atomic limit, or ``Wannierizable", bands correspond to an even parity, whereas an odd parity signals topological bands that cannot be expressed in terms of localized Wannier functions~\cite{vaidya2023topological, ghorashi2024prevalence, benalcazar2019quantization}.

With preserved time-reversal $(\mathcal{T})$ symmetry, the topological bands enforce Dirac points in the interior of the Brillouin zone, protected by the combined $C_{2}\mathcal{T}$ symmetry. These linear degeneracies can give rise to a vanishing density of states when frequency isolated and enable applications such as large-area single-mode lasing~\cite{dirac_laser1, dirac_laser2, BerkSEL} and light confinement~\cite{diracmode1, diracexp1, diracexp2, diracexp3, vaidya2021point}. When $\mathcal{T}$ is broken, the same eigenvalue patterns correspond to bands with nonzero Chern number, which support unidirectional edge states that propagate without backscattering. Such edge channels enable slow light with a large bandwidth and can be used for optical isolation~\cite{wang2008reflection, wang2025topological, Slowlight_chern1, Slowlight_chern2, Slowlight_chern3}. Both manifestations follow from the same symmetry indicators extracted from the $C_{2}$ eigenvalues.

Bands in the atomic limit, both with and without $\mathcal{T}$, support a Wannier representation centered at one of the maximal Wyckoff positions (\cref{figure: 1}(c)) of the lattice and carry quantized bulk polarization and corner charge invariants. These invariants determine whether the system hosts boundary-localized modes at edges or corners, even when the bulk is fully gapped. 

Within both the atomic-limit and topological categories, the four distinct eigenvalue patterns form subcategories that correspond to different weak invariants such as nontrivial polarization components or corner charge values, which enrich the classification~\cite{vaidya2023topological, vaidya2023response, benalcazar2019quantization}. This symmetry-based framework characterizes the band topology for PhCs in the spacegroup $p2$ and motivates the use of $C_{2}$ eigenvalues as a classification target for machine learning and inverse design.

\paragraph{PhC dataset}
To train our machine learning models, we create a dataset consisting of 40,000 two-dimensional PhCs possessing two-fold rotational symmetry, i.e., $\varepsilon(\mathbf{r}) = \varepsilon(-\mathbf{r})$, with band properties computed using MIT Photonic Bands \cite{johnson2001block} and analyzed using MPBUtils.jl~\cite{MPBUtils.jl} and Crystalline.jl~\cite{Crystalline.jl}, with post-processing as detailed in our publically available Github repository \cite{Sparse-Networks-For-Topological-Photonics}. Our approach represents the dielectric function as Fourier components instead of image representations as input parameters. This choice is motivated by two key factors. First, it significantly improves interpretability: a model based on a small number of Fourier components is far easier to analyze than one which uses 1,024 image pixels (e.g., a $32 \times 32$ input grid). Second, from a physical standpoint, arguments from perturbation theory (SM Section S3) suggest that the primary physical information necessary for the analysis of both band dispersion and band symmetry is contained in this  representation, despite its dimensional sparsity~\cite{kittel2018introduction, kaxiras2019quantum}.

The dielectric function of each PhC, $\varepsilon(\mathbf{r})$, represents the spatial variation of the dielectric constant in each unit cell. This is a continuous function constructed through random sampling of the Fourier components corresponding to the smallest 19 reciprocal lattice vectors~\footnote{Our ordering of reciprocal lattice vectors is consistent with \cite{Crystalline.jl}. In particular, we order from smallest to largest norms (in the basis of primitive reciprocal lattice vectors).}. Explicitly, 
\begin{equation}
\varepsilon(\mathbf{r}) = \sum_{\mathbf{G} \neq \bm{0}} \varepsilon_\mathbf{G} e^{i\mathbf{G} \cdot \mathbf{r}},
\end{equation}
where the Fourier components $\varepsilon_\mathbf{G}$ (for $\mathbf{G} \neq \mathbf{0}$) are independently sampled from a uniform distribution between -1 and 1. To ensure physically meaningful dielectric profiles (i.e., $\varepsilon(\mathbf{r}) > 1$ everywhere), we fix the static frequency component $\varepsilon_{\bm{0}} = 20$. Additionally, enforcing two-fold rotational symmetry requires $\varepsilon_\mathbf{G} = \varepsilon_{-\mathbf{G}}$, reducing the number of independent components to nine. Although the PhCs in our training dataset have smooth dielectric profiles and small dielectric contrasts close to unity (\cref{figure: 1}(a)), we show later that our inverse-design pipeline generalizes and extends to experimentally realizable two-tone dielectric profiles with arbitrary dielectric contrast. For simplicity, we also fix the lattice vectors of our PhCs to $\mathbf{R}_1 = a(1, 0)$ and $\mathbf{R}_2 = a(1/2, 4/5)$ in Cartesian coordinates (where the overall length scale, $a$, is irrelevant to topological properties of the bands). Note that our choice of lattice vectors is arbitrary, and our methods are agnostic to the particular choice of lattice vectors \footnote{One could also vary the lattice vectors across the dataset though this would lead to more complicated models}.

Additionally, we augment the dataset by exploiting the symmetry of the underlying plane group: each PhC is mapped to three related PhCs obtained by translating the unit cell by half a lattice vector along either basis direction (\cref{figure: 1}(a)). These shifts relocate the unit cell origin to one of three other maximal Wyckoff positions~\cite{aroyo2006bilbao} (\cref{figure: 1}(c)). Importantly, this transformation preserves the $C_2$ symmetry and modifies the $C_2$ eigenvalues at the high-symmetry points deterministically. The corresponding mapping between Fourier components is also analytically known. As a result, using data augmentation rooted in topological quantum chemistry~\cite{kruthoff2017topological, bradlyn2017topological, po2017symmetry}, we expand our dataset to 160,000 unique PhCs, without need for additional simulations. For concreteness, the explicit forms of these mappings are given below \cite{ghorashi2024prevalence}.
\begin{equation}
\begin{split}
\varepsilon'_\mathbf{G}=\varepsilon_\mathbf{G}\mathrm{e}^{\mathrm{i}\mathbf{G}\cdot\mathbf{\Delta}}\\
    x'_\mathbf{k} = x_\mathbf{k} \mathrm{e}^{2\mathrm{i}\mathbf{k}\cdot\mathbf{\Delta}},
    \label{equation; data augmentation}
    \end{split}
\end{equation}
where $\mathbf{\Delta}$ is the unit cell shift, $x_\mathbf{k}$ is the $C_2$ symmetry eigenvalue at point $\mathbf{k}$ (for a given band), and we denote with primes quantities corresponding to the shifted PhC. 

We investigate the lowest-lying transverse-magnetic (TM) polarized modes in detail here and leave the transverse-electric (TE) polarized modes and higher-lying bands to Section S2 of the Supplemental Materials (SM). 
\begin{figure}
    \centerline{%
    \includegraphics[scale=1]{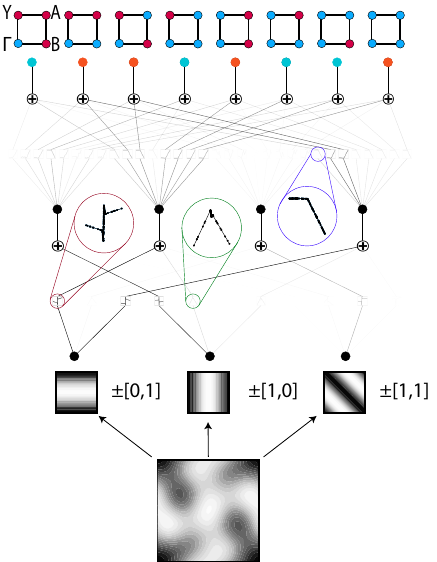}%
    }
    \caption{%
        Kolmogorov--Arnold network trained on $C_2$ symmetric PhCs. The pruned network only depends on three input parameters, with only four nodes in the hidden layer. Three trained activation functions are highlighted in red, green, and blue and shown in more detail in the insets. Points lying on top of the activation functions denote samples from the dataset. We show the decomposition of the dielectric function of the PhC from \cref{figure: 1} into the lowest three Fourier components, which comprise the three input parameters. The symmetry classes are labeled by the same convention as in \cref{figure: 1}(a). Classes with an odd (even) number of red circles correspond to topological (trivial) bands. We show the trivial (topological) classes in orange (turquoise). 
        }
        \label{figure: 2}
\end{figure}

\textbf{Kolmogorov--Arnold networks for photonic topology.} Kolmogorov–Arnold Networks (KANs)~\cite{liu2024kan} are a class of neural networks inspired by the Kolmogorov–Arnold representation theorem, in contrast to traditional multilayer perceptrons (MLPs) which draw from the universal approximation theorem. KANs differ from MLPs by assigning learnable one-dimensional functions, parametrized as splines, to the edges rather than fixed activation functions to the nodes. This structure replaces the conventional linear weight matrices with nonlinear, trainable activation functions. KANs are further regularized through sparsity-promoting training and pruning, resulting in compact models that lend themselves to symbolic regression. These features make it possible to extract interpretable, algebraic relationships between inputs and outputs, making KANs particularly suited for discovering underlying physical laws from scientific data.

We train a KAN classifier on our data-augmented PhC dataset where the Fourier components of the dielectric function serve as input features for predicting the topological class of the lowest TM band. On training with sparsity, the model retains only three of the nineteen input parameters while reaching $>99\%$ accuracy (\cref{figure: 2}). This indicates that the network identifies the smallest three Fourier coefficients (those corresponding to the smallest reciprocal lattice vectors) as the decisive parameters governing the topology of the lowest band, consistent with perturbation theory arguments (SM Section S3). While this strong performance is obtained on a large dataset of 160,000 PhCs, we find that the same architecture also performs exceedingly well in data-scarce settings, especially when data augmentation is used. In SM Section S4, we show that a training set of only 64 PhCs reaches nearly $90\%$ accuracy with augmentation, compared to $\sim 50\%$ without it.

Next, we use symbolic regression \cite{liu2024kan} to map the neural network to eight formulas, $f_{1}, f_{2},...,f_{8}$, one for each topological class, which we report in the SM (SM Section S5). Since symbolic regression is performed after the pruning of the network, each function, $f_n$, is only a function of three variables: $\varepsilon_{01}$, $\varepsilon_{10}$ and $\varepsilon_{11}$, where we have introduced $\varepsilon_{ml}\equiv\varepsilon_{m\mathbf{G}_1+l\mathbf{G}_2}$. Each formula, $f_n$, corresponds to the likelihood of the corresponding symmetry class, $n$. Concretely, to predict the symmetry class of a given photonic crystal, we calculate: 
\begin{equation}
\begin{split}
        \text{argmax}\{f_1(\varepsilon_{10}, \varepsilon_{0 1}, \varepsilon_{11}), f_2(\varepsilon_{10}, \varepsilon_{0 1}, \varepsilon_{11}),...,\\ f_8(\varepsilon_{10}, \varepsilon_{01}, \varepsilon_{11})  \}
        \end{split}
        \label{Equation: formula classification}
\end{equation}
To motivate the physical interpretation of our formulas, we show in the SM (Section S5) that the topological classification obtained via \cref{Equation: formula classification} is equivalent to predictions from perturbation theory. This observation shows that the KAN model effectively reconstructs the perturbative mechanism by which the lowest Fourier components control the splitting of symmetry representations at high-symmetry momenta.

\begin{figure}
    \includegraphics[scale=1]{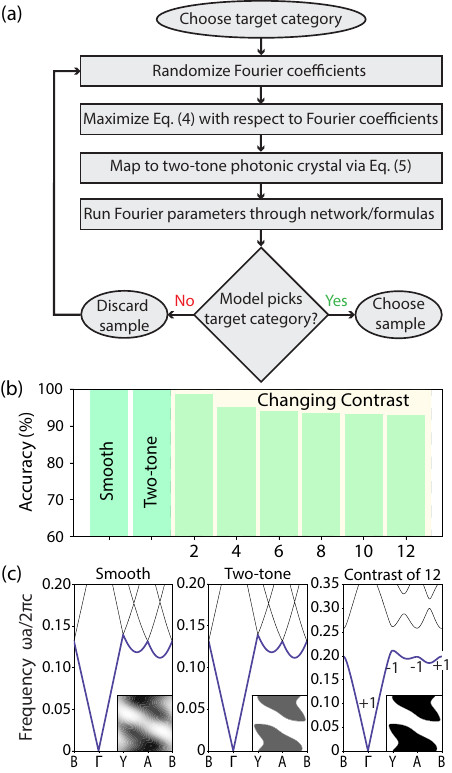}
    \caption{%
        (a) Inverse design workflow. We first choose a target category and pick a random PhC to serve as a starting point. We next do gradient descent on the PhC's Fourier components to yield a smooth lattice within the target category. We then used \cref{equation: two-tone-mapping} to map our smooth PhCs to two-tone PhCs. 
        (b) Accuracy of our inverse design scheme. Blue (yellow) bars correspond to inverse design from the KAN (formulas), respectively. We achieve greater than 99 percent accuracy in creating smooth lattices and low-contrast two-tone lattices. Increasing the dielectric contrast up to 12 lowers the accuracy only moderately, to 92 percent. (c) Band structures at each stage of the inverse design workflow. $C_2$ Symmetry eigenvalues for the lowest band remain unchanged throughout the workflow and are shown overlaid on the rightmost band structure.  
        }
        \label{figure: 3}
\end{figure}

\textbf{Inverse design.} To generate PhCs with desired topological class for the lowest band, we use the methodology outlined in \cref{figure: 3}(a). First, we pick a random PhC as a starting point generated by sampling the Fourier coefficients from a uniform distribution. Because the trained model identifies only the smallest three Fourier coefficients as relevant for the band topology, the remaining coefficients serve only to diversify the unit-cell geometry without influencing the resulting topological class.

Next, we define the following objective function for optimization based on the formulas obtained from the trained KAN model:
\begin{equation}
    C_\text{T} f_n(\vec{\varepsilon})-\sum_{m\neq n} f_m(\vec{\varepsilon}),
    \label{equation: inverse design}
\end{equation}
where $n$ is the target topological class, $C_\text{T}$ is the number of topological classes and we denote by $\vec{\varepsilon}$ the vector of Fourier components. Optimizing \cref{equation: inverse design} corresponds to increasing the liklihood that the PhC design belongs to the target class while minimizing the likelihood of other classes. We then perform finite-difference gradient descent on the three input parameters using this objective function. In all, we perform this process on 2000 PhCs targeting each of the eight topological classes of the lowest TM band. 

Experimentally realizable PhCs are often ``two-tone", \ie consisting of only two materials with distinct refractive indices. To map these generated smooth PhCs to experimentally realizable two-tone PhCs, we define: 
\begin{equation}
    \varepsilon_{\text{two-tone}}(\mathbf{r})=\Theta(\varepsilon(\mathbf{r})-\varepsilon_{\text{50\%}})(\varepsilon_{75\%}-\varepsilon_{25\%})+\varepsilon_{25\%},
    \label{equation: two-tone-mapping}
\end{equation}
where $\varepsilon(\mathbf{r})$ is the smooth dielectric function, $\varepsilon_{\text{two-tone}}(\mathbf{r})$ is the two-tone dielectric function, $\Theta$ is the Heaviside step function and $\varepsilon_\text{x\%}$ denotes the $\textit{x}$'th percentile of the smooth dielectric function throughout the unit cell. We then increase the dielectric contrast of these two-tone lattices, setting the lower contrast region of the two-tone unit cells to vacuum ($\varepsilon=1$) and the higher contrast region to values ranging from 2 to 12. 

It is critical to note that the thresholding operation defined in Eq. \ref{equation: two-tone-mapping} constitutes a substantial modification to the continuous dielectric function, which changes its Fourier components and could potentially alter the topological class of the band. To ensure that the two-tone PhC continues to belong to the target topological class, we utilize our KAN-derived analytic formulas as a fast ``surrogate solver" to verify the topological classification. If a deviation from the target is found, we discard the sample and restart the optimization with another random initialization. 

\begin{figure}
    \includegraphics[scale=1]{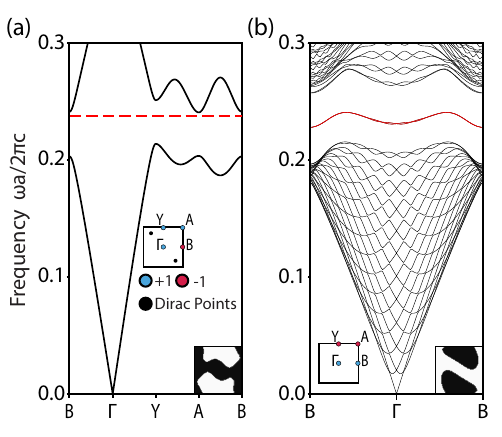}
    \caption{Inverse designed Dirac points and edge states. (a) Band structure corresponding to one of the inverse designed PhCs with nontrivial topology below the first gap. Dashed red line denotes the frequency of the Dirac degeneracy, which occurs in the interior of the Brillouin zone. (b) Band structure corresponding to a one-dimensional tiling of an atomic limit. The edge state, denoted by red, arises from the difference in polarization between the bulk and the trivial cladding of the tiling. Here, the cladding is a shifted version of the same unit cell, obtained using \cref{equation; data augmentation}.
        }
        \label{figure: 4}
\end{figure}

Using this workflow, we find that our inverse design method maintains an accuracy $> 92\%$ (\cref{figure: 3}(b)), for generating two-tone topological PhCs across a range of dielectric contrasts. This accuracy indicates that the learned formulas, which capture aspects of degenerate perturbation theory in the small-contrast regime, generalize beyond the training domain and continue to predict the topology of the lowest band at higher contrasts (\cref{figure: 3}(c)). For completeness, in the SM (Section S6), we show the accuracy of our inverse design method for each topological category.

Finally, we demonstrate two examples of experimentally realizable PhCs designed using our inverse design workflow. \cref{figure: 4}(a) shows a structure whose lowest band exhibits topological bands. The dispersion hosts frequency-isolated Dirac points, and the corresponding real-space unit cell produced by the inverse design is shown in the inset. In \cref{figure: 4}(b), we present a structure that hosts an atomic-limit band with non-trivial polarization. The resulting ``slab band structure" contains an in-gap edge mode localized at the interface between the bulk region and a trivial cladding.

\paragraph{Discussion}
We have demonstrated that a Kolmogorov--Arnold network (KAN) with a single hidden layer can predict the topological classes of transverse-magnetic (TM) polarized modes below the fundamental gap with over 99\% accuracy. Using symbolic regression, we have translated the network’s behavior into a concise set of eight analytical formulas, each corresponding to a target class, enabling the inverse design of topological PhCs. Although in this work we also employ post-processing in our inverse design workflow to obtain experimentally realizable two-tone photonic crystals, no tools beyond the trained KAN/formulas themselves are used. 

KANs offer several advantages over other machine learning architectures. As we have observed, the resulting expressions are interpretable and reveal the physical content learned by the network. They identify the smallest Fourier components as the dominant parameters controlling the topology of the lowest bands. They also show, surprisingly, that the topology of these bands is consistent with expectations from perturbation theory, even in high dielectric contrast regimes. Furthermore, we find that KANs are able to perform well in data scarce scenarios, achieving an accuracy of $> 90\%$ with fewer than 100 PhCs for training in our task.

Our findings open several promising directions for future work. A natural extension would be to apply this framework to more complicated two-dimensional plane groups or three-dimensional space groups, particularly those with higher rotational symmetry or additional mirror and non-symmorphic symmetries. Such generalizations would need to address potential challenges related to band connectivity and degeneracies at high-symmetry points, which could either be enforced by-hand or otherwise by extending the classification scheme to account for these additional considerations. 

Beyond photonics, it would also be interesting to explore similar inverse design workflows in predicting structural or electronic properties in materials science. For crystalline materials, a natural analog for the Fourier components of the dielectric function would be the charge density or Kohn--Sham wavefunctions \cite{kohn1965self}, provided in a plane-wave basis. For inverse design of new, experimentally realizable materials, however, ensuring the additional constraint of dynamical and thermodynamic stability \cite{siriwardane2022generative} using density functional theory  \cite{lyngby2022data} is important. Our results suggest that KANs may be well suited for inverse design problems in this related domain, particularly where physical properties are encoded in low-dimensional, symmetry-constrained representations.

\vskip 2ex
\begin{acknowledgments}
A.G., S.V., and M.S. acknowledge support from the U.S.\ Office of Naval Research (ONR) Multidisciplinary University Research Initiative (MURI) under Grant No.\ N00014-20-1-2325 on Robust Photonic Materials with Higher-Order Topological Protection. This material is also supported in part by the Air Force Office of Scientific Research under the award number FA9550-21-1-0317 and in part by the US Army Research Office through the Institute for Soldier Nanotechnologies at MIT, under Collaborative Agreement Number W911NF-23-2-0121. The authors acknowledge support from the National Science Foundation under Cooperative Agreement PHY-2019786 (The NSF AI Institute for Artificial Intelligence and Fundamental Interactions). T.C. acknowledges the support of a research grant (project no. 42106)
from Villum Fonden. The MIT SuperCloud and Lincoln Laboratory Supercomputing Center provided computing resources that contributed to the results and the dataset reported in this work.
\end{acknowledgments}

\bibliographystyle{supp/apsrev4-2-longbib}
\clearpage
\bibliography{main}

\onecolumngrid
\newpage
\renewcommand*{\arraystretch}{1} 

\hyphenation{}
\begin{center}
    {\Large \textbf{SUPPLEMENTAL MATERIAL}}\\[0.75cm]
    {\Large \textbf{Symbolic Learning of Topological Bands in Photonic Crystals}}\\[0.5cm]
    {\large Ali~Ghorashi$^\bigstar$,$^{1, 2, \textasteriskcentered{}}$ Sachin~Vaidya$^\bigstar$,$^{1, 3, 4,}$\textsuperscript{\textdagger} Ziming~Liu,$^{1,4}$
    Charlotte~Loh,$^1$ 
    Thomas~Christensen,$^5$
    Max~Tegmark,$^{1,4}$ Marin~Solja\v{c}i\'c$^{1,3,4}$
    }\\[0.25cm]
    {$^{1}$Department of Physics, Massachusetts Institute of Technology, Cambridge, MA 02139, USA\\
$^{2}$Department of Applied Physics, Yale University, New Haven, Connecticut 06520, USA\\
$^{3}$Research Laboratory of Electronics , Massachusetts Institute of Technology, Cambridge, Massachusetts 02139, USA\\ 
$^{4}$The NSF Institute for Artificial Intelligence and Fundamental Interactions\\
$^{5}$Department of Electrical and Photonics Engineering, Technical University of Denmark, 2800 Kgs.\ Lyngby, Denmark\\}
{$^\bigstar$ denotes equal contribution\\}
{\textasteriskcentered{} aligho@mit.edu   \textdagger svaidya1@mit.edu}
\end{center}

\renewcommand{\theequation}{S\arabic{equation}} 
\renewcommand{\thefigure}{S\arabic{figure}} 
\renewcommand{\thetable}{S\arabic{table}} 
\renewcommand{\thesection}{S\arabic{section}}
\renewcommand{\thesubsection}{\Alph{subsection}}
\renewcommand{\thesubsubsection}{\roman{subsubsection}}



\maketitle

\setlength{\parindent}{0em}
\setlength{\parskip}{.5em}

\section{Summary of supplementary materials}
In this section, we provide a brief overview of these supplementary materials. 
In \cref{section: higher band kans}, we show that Kolmogorov-Arnold networks (KANs) can used to predict band topology beyond the lowest transverse magnetic (TM) mode. In particular, we show that KANs may be trained to predict the topology of the lowest transverse electric (TE) band as well as that of the second TM band.  
In \cref{sec: perturbation theory}, we show that the salient inputs picked by our networks in the main text (i.e. after pruning) are consistent with degenerate perturbation theory. 
In \cref{section: symbolic formulas}, we report the formulas obtained through symbolic regression on the Kolmogorov-Arnold network shown in the main text. These formulas were used for the inverse design process shown in Figure 3 of the main text. In \cref{section: data sparsity}, we show how our data augmentation scheme (Eq. (2) of the main text) can be used to train KANs with very high accuracies in data-scarce scenarios. Finally, in \cref{section: inverse design}, we include details of our inverse design methodology--which we briefly described in the flowchart of Figure 3 in the main text. In addition, we show examples of photonic crystals produced through inverse design and additional inverse design statistics.

\section{KANs for prediction beyond the first TM band}
\label{section: higher band kans}
\begin{figure}
    \centerline{%
    \includegraphics[scale=1]{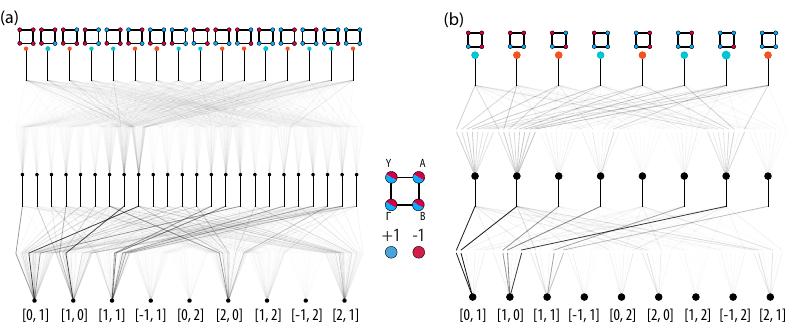}}
    \caption{\textbf{KANs for prediction of band topology beyond the lowest TM band:}
        (a)~KAN for prediction of the band symmetries of the second TM band. (b)~KAN for prediction of the band symmetries of the first transverse-electric polarized band. In each case, Fourier components given in the $\mathbf{G}_1, \mathbf{G}_2$ basis are mapped to the possible symmetry classes, of which there are sixteen for the second TM band and eight for the first TE band. We delineate trivial vs topological classes by labeling the former with orange and the latter with turquoise dots.  
        }
    \label{figure: Kans for other bands}
\end{figure}

Although we focused on the lowest TM band in the main text, KANs can be similarly applied to TE modes and higher bands. To illustrate this point, in \cref{figure: Kans for other bands} we show KANs that were trained to classify the symmetry of the second TM band (\cref{figure: Kans for other bands}(a)) and the first TE band (\cref{figure: Kans for other bands}(b)). For the second TM band, the output targets are 16 different symmetry classes (instead of 8 for the first band), since the second band does not have a pinned symmetry at the $\Gamma$ point. 
We find that KANs achieve a $98\%$ train and $96\%$ test accuracy with only one hidden layer in classifying the second TM band. They achieve similar success in predicting the first TE band symmetry, achieving a $98\%$ train and $98\%$ test accuracy, again with only one hidden layer. 
\FloatBarrier

\section{Topological bands in the perturbative limit}
\label{sec: perturbation theory}
In this section, we obtain predictions for the topological classification of photonic bands in the low contrast limit using degenerate perturbation theory.

We start with Maxwell's equations for nonmagnetic materials ($\mu=\mu_0$) in the absence of sources ($\mathbf{J}=\mathbf{0}, \rho=0$):
\begin{align*}\nabla\times\mathbf{E}(\mathbf{r})=i\omega\mu_0\mathbf{H}(\mathbf{r}), \nabla\times\mathbf{H}(\mathbf{r})=-i\omega\varepsilon_0\varepsilon(\mathbf{r})\mathbf{E}(\mathbf{r}),\end{align*}
When combined, these equations give us the following Hermitian semi-positive-definite eigenvalue problem for the magnetic field, $\mathbf{H}(\mathbf{r})$: 
\begin{align*}\nabla\times\frac{1}{\varepsilon(\mathbf{r})}\nabla\times\mathbf{H}(\mathbf{r})=(\omega/c)^2\mathbf{H}(\mathbf{r})\end{align*}
For a two-dimensional photonic crystal with lattice vectors, $\mathbf{R}_{1}$ and $\mathbf{R}_2$, the permittivity, $\varepsilon(\mathbf{r})$, has discrete translational symmetry: $\varepsilon(\mathbf{r}+n\mathbf{R}_1+m\mathbf{R}_2)=\varepsilon(\mathbf{r})$, for any integers $n$ and $m$. This means that we may generically expand $\varepsilon(\mathbf{r})$ in a Fourier series with quasi-momenta restricted to vectors in the photonic crystal's reciprocal lattice: $\varepsilon(\mathbf{r})=\sum_\mathbf{G}\varepsilon_\mathbf{G}e^{i\mathbf{G}\cdot\mathbf{r}},$ where $\mathbf{G}=n\mathbf{G}_1+m\mathbf{G}_2$ and $\mathbf{G}_i$ are primitive reciprocal lattice vectors which satisfy $\mathbf{R}_j\cdot\mathbf{G}_i=2\pi\delta_{ij}$. To enforce $C_2$  symmetry, i.e. $\varepsilon(\mathbf{r})=\varepsilon(-\mathbf{r})$, we additionally have the constraint:  $\varepsilon_\mathbf{G}=\varepsilon_{-\mathbf{G}}$. For the sake of brevity, in the discussion below, we define $\varepsilon_{nm}\equiv\varepsilon_{\mathbf{G}=n\mathbf{G}_1+m\mathbf{G}_2}$.

For the purposes of symmetry diagnosis, we need only concern ourselves with the dispersion of photonic bands at the high symmetry $\mathbf{k}$-points: $\mathbf{k}=(n\mathbf{G}_1+m\mathbf{G}_2)/2$ (with $(n, m)=(1, 0), (0, 1), (1, 1)$ referred to as the $\textbf{B}$, $\textbf{Y}$, and $\textbf{A}$ points, respectively). At the $\textbf{B}$ point, the two lowest bands in the empty lattice (by which we mean the photonic crystal corresponding to a uniform permittivity $\varepsilon(\mathbf{r})=\varepsilon_{00}$) have wavevectors $\pm \mathbf{G}_1/2$. At the $\textbf{Y}$ point, the two lowest bands have wavevectors $\pm \mathbf{G}_2/2$. However, at the $\textbf{A}$ point, matters are slightly more complicated: if $|\mathbf{G}_1+\mathbf{G}_2|>|\mathbf{G}_1-\mathbf{G}_2|$ (equivalently if $\mathbf{G}_1\cdot\mathbf{G}_2 >0$), the  two lowest empty lattice bands at the $\textbf{A}$ point have wavevectors $\mathbf{k}=\pm(\mathbf{G}_1-\mathbf{G}_2)/2$ and if $\mathbf{G}_1\cdot\mathbf{G}_2 <0$, they have wavevectors $\mathbf{k}=\pm(\mathbf{G}_1+\mathbf{G}_2)/2$. For the photonic crystals used in our dataset, we have $\mathbf{R}_1=a(1, 0)$ and $\mathbf{R}_2=a(1/2, 4/5)$, which means that $\mathbf{G}_1=(\pi/a) (2, -5/4)$ and $\mathbf{G}_2=(2\pi/a)(0, 5/4)$, which means that $\mathbf{G}_1\cdot\mathbf{G}_2<0$. Thus, the symmetries of the two lowest bands at the $\mathbf{A}$ point for our dataset are determined by $\varepsilon_{11}$ and not $\varepsilon_{-11}$.

We note that, in the empty lattice, the two lowest bands are degenerate at the high symmetry points, both having frequencies given by  $\omega_{0, \mathbf{k}}=c|\mathbf{k}|/\sqrt{\varepsilon_{00}}$. In the presence of the perturbation, a gap generically opens up, and we must find the correction to the unperturbed frequency, $\omega_{0, \mathbf{k}}$. To do this, we write the Maxwell eigenproblem  (with $\Delta (\omega_\mathbf{k}/c)^2$ denoting the change in the frequency squared due to the perturbation) as follows: 

\begin{equation}\nabla\times\Bigg[\frac{1}{\sum_\mathbf{G}\varepsilon_\mathbf{G}e^{i\mathbf{G}\cdot\mathbf{r}}}-\frac{1}{\varepsilon_{0,0}} \Bigg]\nabla\times\mathbf{H}(\mathbf{r})\equiv \nabla\times\Bigg[\Delta \frac{1}{\varepsilon(\mathbf{r})} \Bigg]\nabla\times\mathbf{H}(\mathbf{r})=\Delta(\omega_\mathbf{k}/c)^2\mathbf{H}(\mathbf{r}), 
\label{Equation: perturbed eigenproblem}
\end{equation}
where we have introduced the change in the inverse permittivity (i.e. the perturbation): 
\begin{align*}\Delta\frac{1}{\varepsilon(\mathbf{r})}\equiv \frac{1}{\sum_\mathbf{G}\varepsilon_\mathbf{G}e^{i\mathbf{G}\cdot\mathbf{r}}}-\frac{1}{\varepsilon_{00}}\approx-\sum_\mathbf{G}\frac{\varepsilon_{\mathbf{G}}}{\varepsilon_{00}^2}e^{i\mathbf{G}\cdot\mathbf{r}}\end{align*}
In the following two subsections, we solve this problem for transverse magnetic and transverse electric modes, respectively. 
\subsection{Transverse-magnetic modes}
We first focus on transverse magnetic modes (modes with $\mathbf{H}(\mathbf{r})\cdot \mathbf{e}_z=0$ everywhere). We write the two degenerate TM modes at $\mathbf{k}=(n\mathbf{G}_1+m\mathbf{G}_2)/2$ as: 
\begin{align*}\mathbf{H}_1(\mathbf{r})=e^{i(n\mathbf{G}_1+m\mathbf{G}_2)\cdot\mathbf{r}/2} (n\mathbf{a}_2-m\mathbf{a}_1),  \mathbf{H}_2(\mathbf{r})= e^{-i(n\mathbf{G}_1+m\mathbf{G}_2)\cdot\mathbf{r}/2} (n\mathbf{a}_2-m\mathbf{a}_1),
\end{align*}
where the polarization vector $n\mathbf{a}_2-m\mathbf{a}_1$ is determined from the condition that $\nabla\cdot\mathbf{H}(\mathbf{r})=0$ and $\mathbf{H}(\mathbf{r})\cdot \mathbf{e}_z=0$.
We solve the eigenproblem in the basis of these two states, taking the field to be a linear combination of these modes: $\mathbf{H}(\mathbf{r})=c_1\mathbf{H}_1(\mathbf{r})+c_2\mathbf{H}_1(\mathbf{r})$, which gives us: 
\begin{align*}\nabla\times\mathbf{H}(\mathbf{r})=i(n\mathbf{G}_1+m\mathbf{G}_2)/2\times (n\mathbf{a}_2-m\mathbf{a}_1)\Big[c_1e^{i(n\mathbf{G}_1+m\mathbf{G}_2)\cdot\mathbf{r}/2}-c_2e^{-i(n\mathbf{G}_1+m\mathbf{G}_2)\cdot\mathbf{r}/2} \Big]\rightarrow\\\Bigg[\Delta\frac{1}{\varepsilon(\mathbf{r})}\Bigg]\nabla\times\mathbf{H}(\mathbf{r})\approx -i(n\mathbf{G}_1+m\mathbf{G}_2)/2\times(n\mathbf{a}_2-m\mathbf{a}_1)\frac{\varepsilon_{nm}}{\varepsilon^2_{0 0}}\Big[c_1e^{-i(n\mathbf{G}_1+m\mathbf{G}_2)\cdot\mathbf{r}/2}-c_2e^{i(n\mathbf{G}_1+m\mathbf{G}_2)\cdot\mathbf{r}/2}\Big]\end{align*}

Therefore, the left hand side of the eigenvalue equation, \cref{Equation: perturbed eigenproblem}, is given by:
\begin{align*}\frac{\varepsilon_{nm}}{4\varepsilon^2_{0 0}}|n\mathbf{G}_1+m\mathbf{G}_2|^2\Big[c_1e^{-i(n\mathbf{G}_1+m\mathbf{G}_2)\cdot\mathbf{r}/2}+c_2e^{i(n\mathbf{G}_1+m\mathbf{G}_2)\cdot\mathbf{r}/2} \Big](n\mathbf{a}_2-m\mathbf{a}_1)\end{align*}
Noting that the eigenvalue problem must be independently satisfied for each momentum, we obtain two equations:
\begin{align*}[\varepsilon_{nm}/\varepsilon^2_{00}]|(n\mathbf{G}_1+m\mathbf{G}_2)|^2 c_2=4\Delta(\omega_{\mathbf{k}}/c)^2c_1, [\varepsilon_{n m}/\varepsilon^2_{00}]|(n\mathbf{G}_1+m\mathbf{G}_2)|^2 c_1=4\Delta(\omega_\mathbf{k}/c)^2c_2\end{align*}
This may be cast as a $2\times 2$ matrix: 
\begin{align*}[\varepsilon_{nm}/\varepsilon^2_{00}]|n\mathbf{G}_1+m\mathbf{G}_2|^2\begin{bmatrix} 0 && 1\\1 && 0 \end{bmatrix} \begin{bmatrix}c_1 \\ c_2 \end{bmatrix}=4\Delta(\omega_\mathbf{k}/c)^2\begin{bmatrix}c_1 \\ c_2 \end{bmatrix}\end{align*}
From this, we see that if $\varepsilon_{nm}$ is postive, the even mode will be the lower band (note that $(n\mathbf{a}_2-n\mathbf{a}_1) \rightarrow -(n\mathbf{a}_2-n\mathbf{a}_1)$ under $C_2$). On the other hand, if $\varepsilon_{nm}$ is negative, the odd mode will be the lower band. We summarize this result succinctly for all high symmetry points in \cref{figure: perturbation theory}.

\subsection{TE modes} 
We can extend the discussion to the TE modes at a high symmetry point $\mathbf{k}=(n\mathbf{G}_1+m\mathbf{G}_2)/2$, for which we have \begin{align*}\mathbf{H}(\mathbf{r})=\Big[c_1e^{i(n\mathbf{G}_1+m\mathbf{G}_2)\cdot\mathbf{r}/2}+c_2e^{-i(n\mathbf{G}_1+m\mathbf{G}_2)\cdot\mathbf{r}/2} \Big]\mathbf{e}_z\rightarrow \\ \nabla\times \Bigg[\Delta\frac{1}{\varepsilon(\mathbf{r})} \nabla\times\mathbf{H}(\mathbf{r})\Bigg] \approx  \Bigg[\frac{\varepsilon_{nm}}{\varepsilon^2_{00}}\Bigg]|(n\mathbf{G}_1+m\mathbf{G}_2)/2|^2\times\Big[c_1e^{-i(n\mathbf{G}_1+m\mathbf{G}_2)\cdot\mathbf{r}/2}+c_2e^{i(n\mathbf{G}_1+m\mathbf{G}_2)\cdot\mathbf{r}/2} \Big]\mathbf{e}_z\end{align*}

Note that since $\mathbf{e}_z\rightarrow \mathbf{e}_z$ under $C_2$, the results are flipped from the TM case (e.g. $\varepsilon_{nm}$ positive makes the lower band odd). As with the TM case, we summarize the perturbation theory results for TE modes in \cref{figure: perturbation theory}.

\subsection{Bands above the fundamental gap}

Note that the arguments above may be readily generalized beyond just the first band. For the second band, for instance, we know immediately from the two preceding sections that the symmetry eigenvalues at $\textbf{A}$, $\textbf{B}$ and $\textbf{Y}$ will be determined by $\varepsilon_{01}, \varepsilon_{10}$ and $\varepsilon_{11}$. This is because if the first band is even (odd) at any of these points, the second band will be odd (even), respectively (at least in the perturbative limit). At the $\Gamma$ point, the second and third bands will correspond to plane waves with $\mathbf{k}=\pm \mathbf{G}_1$ and $\mathbf{k}=\pm (\mathbf{G}_1+\mathbf{G}_2)$. $\mathbf{G}_1$ and $\mathbf{G}_1+\mathbf{G}_2$ will be coupled via $\varepsilon_{01}$, $\mathbf{G}_1$ and $-\mathbf{G}_1-\mathbf{G}_2$ will be coupled via $\varepsilon_{21}$, and $\mathbf{G}_1$ and $-\mathbf{G}_1$ will be coupled by $\varepsilon_{20}$. These correspond exactly with the most dominant input parameters deduced by the KAN shown in \cref{figure: Kans for other bands}.

\subsection{Comparison to derivation from \cite{joannopoulos2008molding}}
The results above may also be obtained following \cite{joannopoulos2008molding}. Our starting point is Eq. (28) in Chapter 2, which we reiterate using the notation introduced above: 
\begin{align*}
\omega_\mathbf{k}-\omega_{0, \mathbf{k}}=-\frac{\omega_{0, \mathbf{k}} \varepsilon_{nm}}{2\varepsilon_{00}}\frac{\int_\Omega e^{i(n\mathbf{G}_1+m\mathbf{G}_2)\cdot\mathbf{r}}|\mathbf{E}(\mathbf{r})|^2\text{d}\Omega}{\int_\Omega |\mathbf{E}(\mathbf{r})|^2\text{d}\Omega},
\end{align*}
where $\omega_\mathbf{k}-\omega_{0, \mathbf{k}}$ is the change in the frequency of a given mode and the squared magnitude of the electric field is given by (up to an unimportant proportionality factor):
\begin{align*}|\mathbf{E}(\mathbf{r})|^2\propto e^{i(n\mathbf{G}_1+m\mathbf{G}_2)}\pm 2+e^{-i(n\mathbf{G}_1+m\mathbf{G}_2)},
 \end{align*}
 where the plus (minus) sign corresponds to the electric field varying as cosine (sine). The integrals may be readily evaluated, since the volume of integration spans the unit cell, $\Omega$:
\begin{align*}\omega_{\mathbf{k}}-\omega_{0, \mathbf{k}}=-\frac{\omega_{0, \mathbf{k}}}{2}\frac{\varepsilon_{nm}}{\varepsilon_{00}}\frac{\int\Big[e^{i(n\mathbf{G}_1+m\mathbf{G}_2)\cdot\mathbf{r}}+c.c\Big]\Big[e^{i(n\mathbf{G}_1+m\mathbf{G}_2)\cdot\mathbf{r}}+c.c \Big]}{\pm 2\Omega}=\mp\frac{\omega_{0, \mathbf{k}}}{2}\frac{\varepsilon_{nm}}{\varepsilon_{00}} \end{align*}
This gives us: 
\begin{align*}\Delta(\omega_\mathbf{k}/c)^2=\frac{\omega_{0, \mathbf{k}}^2}{c^2}\Bigg[1\mp\frac{\varepsilon_{nm}}{\varepsilon_{00}} \Bigg]-\frac{\omega_{0,\mathbf{k}}^2}{c^2}=\mp \frac{|n\mathbf{G}_1+m\mathbf{G}_2|^2\varepsilon_{nm}}{4\varepsilon_{00}^2},\end{align*}
where we have neglected terms of order $(\varepsilon_{nm}/\varepsilon_{00})^2$ and used the expression for the unperturbed frequency: 
$\omega_{0,\mathbf{k}}^2/c^2=|n\mathbf{G}_1+m\mathbf{G}_2|^2/4\varepsilon_{00}$.
In the above, the upper sign corresponds to the cosine  mode (for the electric field--not magnetic field; i.e. even for TM and odd for TE) and the lower sign corresponds to the sine mode (odd for TM and even for TE). Therefore, as in our previous derivation, we see that $\varepsilon_{nm}$ positive means the first TM band will be even under $C_2$ and the first TE band will be odd under $C_2$. 

\begin{figure}
    \centerline{%
    \includegraphics[scale=1]{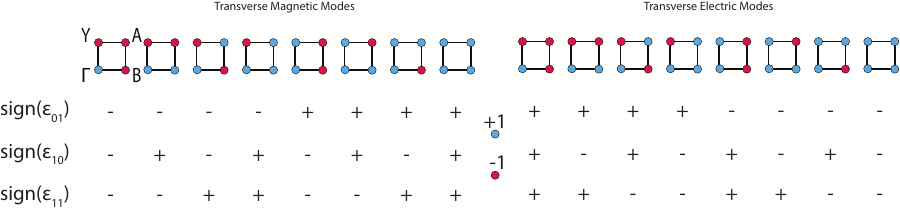}}
\caption{\textbf{Perturbation theory prediction for the lowest TM and TE modes:} Columns in each table correspond to the distinct topological classes, and rows correspond to the signs of the three salient Fourier components of the dielectric function.}
    \label{figure: perturbation theory}
\end{figure}

\section{Data sparsity}
\label{section: data sparsity}
\begin{figure}[hb]
    \centerline{%
    \includegraphics[scale=1]{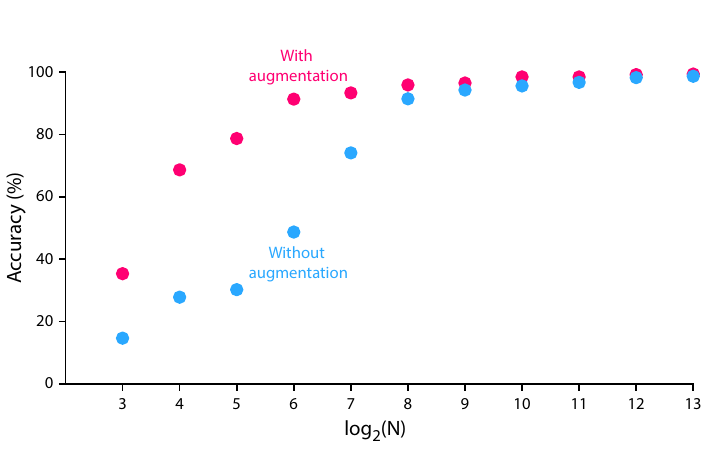}}
\caption{\textbf{Usefulness of augmentation for small data sets:} We compare the accuracy of the KAN (on the entire dataset) after it has been trained on only a small number training data. Light blue dots correspond to accuracies after being trained on $N$ data points, whereas red data points correspond to accuracies after the KAN has been trained on the same $N$ data points but with four possible augmentations (meaning the training data has been expanded to $4N$ samples). 
        }
    \label{figure: augmentation usefulness}
\end{figure}

In the main text, we employed data augmentation using methods from Topological Quantum Chemistry (TQC)~\cite{bradlyn2017topological}. Here, we highlight one of the main benefits of using such a physics-informed augmentation method: the ability to train our networks in a data-scarce setting. In \cref{figure: augmentation usefulness}, we show the performance of KANs trained on PhC datasets with very few samples (ranging from $8$ samples in the smallest dataset to $8192$ samples in the largest dataset). We show results with (red data points) and without (blue data points) augmentation. We find that augmentation drastically improves accuracy for datasets with fewer than a few hundred samples, reaching $\sim 90\%$ accuracy with only $64$ samples. Surprisingly, we also find that KAN trained on $4N$ random photonic crystals performs significantly worse than a KAN trained on an augmented dataset with $4N$ PhCs (an original dataset size of $N$ before augmentation). We reason that this is because the augmented dataset is derived from physical laws and group-theoretic considerations, which are only apparent for a random sampling of photonic crystals when the total number of samples is large. 

\section{Symbolic formulas for the lowest TM band}
\label{section: symbolic formulas}

As stated in the main text, KANs may be mapped to equivalent symbolic formulas. Such formulas may then be used for inverse design, as they allow for construction of an analytic objective function. 

Here, we report the symbolic formulas corresponding to classification of the lowest TM band's symmetry. We use the notation $f_{n}(x_1, x_2, x_3)$ for the formulas, with $n\in\{1, 2,...8\}$ labeling the topological class and $x_1\equiv \varepsilon_{01}, x_2\equiv \varepsilon_{10}$ and $x_3\equiv \varepsilon_{11}$ corresponding to the salient Fourier components of the dielectric function. As stated in the main text, to classify the topological class, we calculate: 
\begin{equation}
    \text{argmax}\left(f_1,f_2,...,f_8 \right),
\end{equation}
with the formulas explicitly given by:
\small
\begin{multline}
    f_1 = -129.609 \sin\left(37.858 \left(- 0.009 x_{1} - 1\right)^{2} + 0.269 \left(- 0.14 x_{2} - 1\right)^{2} - 45.538 + 0.03 e^{- 1.591 \left(1 - 0.379 x_{3}\right)^{2}} \right) \\ +2969.801 \cos{\left(0.773 x_{2} - 0.04 \left(1 - 0.049 x_{3}\right)^{2} + 0.005 \left(- 0.953 x_{1} - 1\right)^{2} - 9.719 \right)} -\\ 2658.012 \cos{\left(0.011 \left(1 - 0.035 x_{2}\right)^{2} + 0.001 \sin{\left(0.909 x_{1} + 7.796 \right)} - 11.023 \sin{\left(0.055 x_{3} + 9.374 \right)} + 10.277 \right)} - \\2359.975 \cos{\left(8.398 \sin{\left(0.027 x_{1} + 9.295 \right)} + 0.005 \tanh{\left(0.482 x_{3} - 0.682 \right)} - 10.845 + 0.003 e^{- 1.608 \left(1 - 0.393 x_{2}\right)^{2}} \right)} - 2084.424 
\end{multline}
\begin{multline}
    f_2 = 54.055 \sin{\left(65.227 \left(- 0.009 x_{1} - 1\right)^{2} + 0.464 \left(- 0.14 x_{2} - 1\right)^{2} - 73.133 + 0.051 e^{- 1.591 \left(1 - 0.379 x_{3}\right)^{2}} \right)} - \\2813.432 \cos{\left(0.385 x_{2} - 0.02 \left(1 - 0.049 x_{3}\right)^{2} + 0.003 \left(- 0.953 x_{1} - 1\right)^{2} - 9.767 \right)} -\\ 2273.723 \cos{\left(0.011 \left(1 - 0.035 x_{2}\right)^{2} + 0.001 \sin{\left(0.909 x_{1} + 7.796 \right)} - 11.214 \sin{\left(0.055 x_{3} + 9.374 \right)} + 10.311 \right)} - \\2401.895 \cos{\left(13.949 \sin{\left(0.027 x_{1} + 9.295 \right)} + 0.008 \tanh{\left(0.482 x_{3} - 0.682 \right)} - 11.547 + 0.004 e^{- 1.608 \left(1 - 0.393 x_{2}\right)^{2}} \right)} - 7030.143 
\end{multline}
\begin{multline}
    f_3 = 87.434 \sin{\left(47.541 \left(- 0.009 x_{1} - 1\right)^{2} + 0.338 \left(- 0.14 x_{2} - 1\right)^{2} - 55.287 + 0.037 e^{- 1.591 \left(1 - 0.379 x_{3}\right)^{2}} \right)} + \\2979.783 \cos{\left(0.771 x_{2} - 0.04 \left(1 - 0.049 x_{3}\right)^{2} + 0.005 \left(- 0.953 x_{1} - 1\right)^{2} - 9.727 \right)} + \\2547.726 \cos{\left(0.01 \left(1 - 0.035 x_{2}\right)^{2} + 0.001 \sin{\left(0.909 x_{1} + 7.796 \right)} - 10.371 \sin{\left(0.055 x_{3} + 9.374 \right)} + 10.238 \right)} - \\2579.134 \cos{\left(10.938 \sin{\left(0.027 x_{1} + 9.295 \right)} + 0.006 \tanh{\left(0.482 x_{3} - 0.682 \right)} - 11.166 + 0.003 e^{- 1.608 \left(1 - 0.393 x_{2}\right)^{2}} \right)} + 2877.984 
\end{multline}
\begin{multline}
    f_4 = - 156.052 \sin{\left(45.795 \left(- 0.009 x_{1} - 1\right)^{2} + 0.326 \left(- 0.14 x_{2} - 1\right)^{2} - 53.509 + 0.036 e^{- 1.591 \left(1 - 0.379 x_{3}\right)^{2}} \right)} - \\3041.197 \cos{\left(0.4 x_{2} - 0.021 \left(1 - 0.049 x_{3}\right)^{2} + 0.003 \left(- 0.953 x_{1} - 1\right)^{2} - 9.742 \right)} + \\2515.52 \cos{\left(0.01 \left(1 - 0.035 x_{2}\right)^{2} + 0.001 \sin{\left(0.909 x_{1} + 7.796 \right)} - 9.965 \sin{\left(0.055 x_{3} + 9.374 \right)} + 10.231 \right)} - \\2289.262 \cos{\left(9.999 \sin{\left(0.027 x_{1} + 9.295 \right)} + 0.005 \tanh{\left(0.482 x_{3} - 0.682 \right)} - 11.029 + 0.003 e^{- 1.608 \left(1 - 0.393 x_{2}\right)^{2}} \right)} - 2801.625 
\end{multline}
\begin{multline}
    f_5 = 46.408 \sin{\left(48.052 \left(- 0.009 x_{1} - 1\right)^{2} + 0.342 \left(- 0.14 x_{2} - 1\right)^{2} - 55.806 + 0.037 e^{- 1.591 \left(1 - 0.379 x_{3}\right)^{2}} \right)} + \\2965.158 \cos{\left(0.772 x_{2} - 0.04 \left(1 - 0.049 x_{3}\right)^{2} + 0.005 \left(- 0.953 x_{1} - 1\right)^{2} - 9.732 \right)} \\- 2450.064 \cos{\left(0.011 \left(1 - 0.035 x_{2}\right)^{2} + 0.001 \sin{\left(0.909 x_{1} + 7.796 \right)} - 11.005 \sin{\left(0.055 x_{3} + 9.374 \right)} + 10.283 \right)} +\\ 2387.297 \cos{\left(26.917 \sin{\left(0.027 x_{1} + 9.295 \right)} + 0.015 \tanh{\left(0.482 x_{3} - 0.682 \right)} - 13.192 + 0.008 e^{- 1.608 \left(1 - 0.393 x_{2}\right)^{2}} \right)} + 2779.054 
\end{multline}
\begin{multline}
    f_6 = - 21.522 \sin{\left(55.237 \left(- 0.009 x_{1} - 1\right)^{2} + 0.393 \left(- 0.14 x_{2} - 1\right)^{2} - 63.037 + 0.043 e^{- 1.591 \left(1 - 0.379 x_{3}\right)^{2}} \right)} - \\3049.487 \cos{\left(0.402 x_{2} - 0.021 \left(1 - 0.049 x_{3}\right)^{2} + 0.003 \left(- 0.953 x_{1} - 1\right)^{2} - 9.738 \right)} - \\2668.682 \cos{\left(0.01 \left(1 - 0.035 x_{2}\right)^{2} + 0.001 \sin{\left(0.909 x_{1} + 7.796 \right)} - 10.39 \sin{\left(0.055 x_{3} + 9.374 \right)} + 10.226 \right)} + \\2499.521 \cos{\left(26.081 \sin{\left(0.027 x_{1} + 9.295 \right)} + 0.014 \tanh{\left(0.482 x_{3} - 0.682 \right)} - 13.096 + 0.008 e^{- 1.608 \left(1 - 0.393 x_{2}\right)^{2}} \right)} - 3089.361 
\end{multline}
\begin{multline}
    f_7 = - 10.691 \sin{\left(69.072 \left(- 0.009 x_{1} - 1\right)^{2} + 0.491 \left(- 0.14 x_{2} - 1\right)^{2} - 76.992 + 0.054 e^{- 1.591 \left(1 - 0.379 x_{3}\right)^{2}} \right)}\\ + 2913.552 \cos{\left(0.773 x_{2} - 0.04 \left(1 - 0.049 x_{3}\right)^{2} + 0.005 \left(- 0.953 x_{1} - 1\right)^{2} - 9.735 \right)} +\\ 2486.685 \cos{\left(0.01 \left(1 - 0.035 x_{2}\right)^{2} + 0.001 \sin{\left(0.909 x_{1} + 7.796 \right)} - 10.403 \sin{\left(0.055 x_{3} + 9.374 \right)} + 10.242 \right)} +\\ 2478.788 \cos{\left(26.99 \sin{\left(0.027 x_{1} + 9.295 \right)} + 0.015 \tanh{\left(0.482 x_{3} - 0.682 \right)} - 13.198 + 0.008 e^{- 1.608 \left(1 - 0.393 x_{2}\right)^{2}} \right)} + 7480.378 
\end{multline}
\begin{multline}
   f_8 = 124.41 \sin{\left(51.096 \left(- 0.009 x_{1} - 1\right)^{2} + 0.363 \left(- 0.14 x_{2} - 1\right)^{2} - 58.852 + 0.04 e^{- 1.591 \left(1 - 0.379 x_{3}\right)^{2}} \right)} - \\3084.661 \cos{\left(0.397 x_{2} - 0.02 \left(1 - 0.049 x_{3}\right)^{2} + 0.003 \left(- 0.953 x_{1} - 1\right)^{2} - 9.733 \right)} + \\2543.803 \cos{\left(0.01 \left(1 - 0.035 x_{2}\right)^{2} + 0.001 \sin{\left(0.909 x_{1} + 7.796 \right)} - 10.048 \sin{\left(0.055 x_{3} + 9.374 \right)} + 10.244 \right)} + \\2306.637 \cos{\left(25.66 \sin{\left(0.027 x_{1} + 9.295 \right)} + 0.014 \tanh{\left(0.482 x_{3} - 0.682 \right)} - 13.043 + 0.008 e^{- 1.608 \left(1 - 0.393 x_{2}\right)^{2}} \right)} + 1799.073 
\end{multline}
\normalsize
\begin{figure}[hb]
\centerline{\includegraphics[scale=0.75]{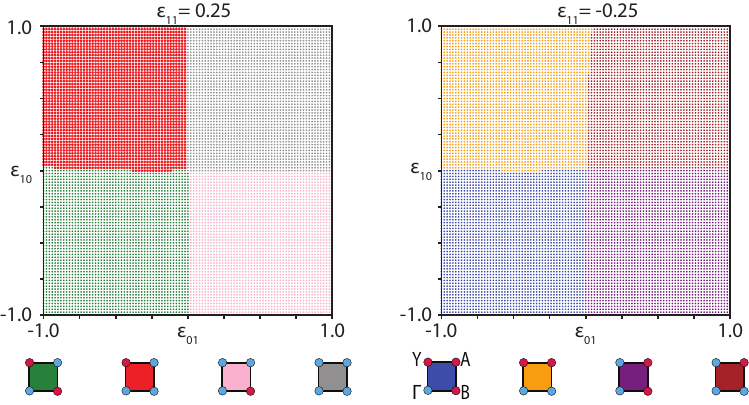}}
\caption{\textbf{Agreement of symbolic formulas with degenerate perturbation theory:} For two slices in ($\varepsilon_{01}, \varepsilon_{10}, \varepsilon_{11}$) space, we show how the prediction of our formulas for the the lowest TM band. The color coded eight categories are shown on the bottom, with their attendant symmetry eigenvalues. In accordance with \cref{figure: perturbation theory}, the formulas predict primarily from the signs of the Fourier components, neatly dividing each slice into four quadrants.}
\label{figure: Formula slices}
\end{figure}

Although the similarity between these formulas and the predictions from perturbation theory given in \cref{figure: perturbation theory} is not immediately apparent, the formulas do, in fact, agree quite well with degenerate perturbation theory. To show this, we plot the topological classifications obtained by these formulas on two slices in three-dimensional $(\varepsilon_{01}, \varepsilon_{10}, \varepsilon_{11})$ space in \cref{figure: Formula slices}. As is apparent, the formulas classify the topological phases entirely based on the sign of the Fourier coefficients, in agreement with the theoretical predictions shown in \cref{figure: perturbation theory}.

\section{Inverse design}
\label{section: inverse design}

\begin{figure}[h]
    \centerline{%
    \includegraphics[scale=1]{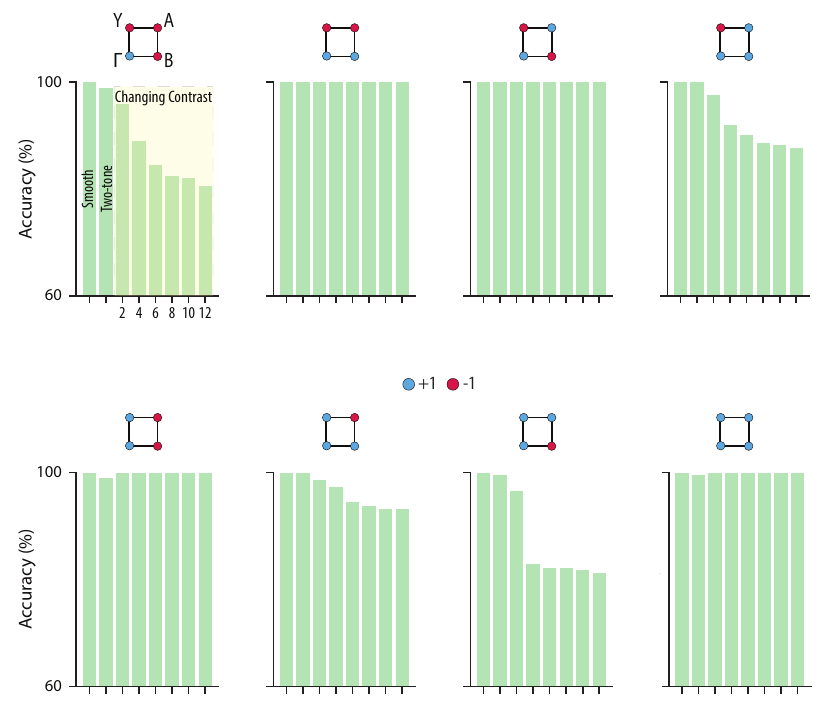}}
    \caption{\textbf{Inverse design accuracies for each topological category:} We delineate the accuracies shown in Figure 3 of the main text by topological category. First two bars in each graph correspond to smooth and low-contrast two tone samples, respectively. The following six bars correspond to accuracies as contrast is increased to experimentally viable values. Note that our inverse design method retains close to 100\% accuracy in designing PhCs that belong to trivial classes.
        }
    \label{figure: inverse design accuracies per class}
\end{figure}

In Figure 3 of  main text, we showed the accuracy of our inverse design process across all topological classes. Here, we show the accuracy delineated by topological class in \cref{figure: inverse design accuracies per class}. Interestingly, we find that our inverse design method remains highly accurate across all topological classes in the high-contrast regime. 

\begin{figure}[]
    \centerline{%
    \includegraphics[scale=0.9]{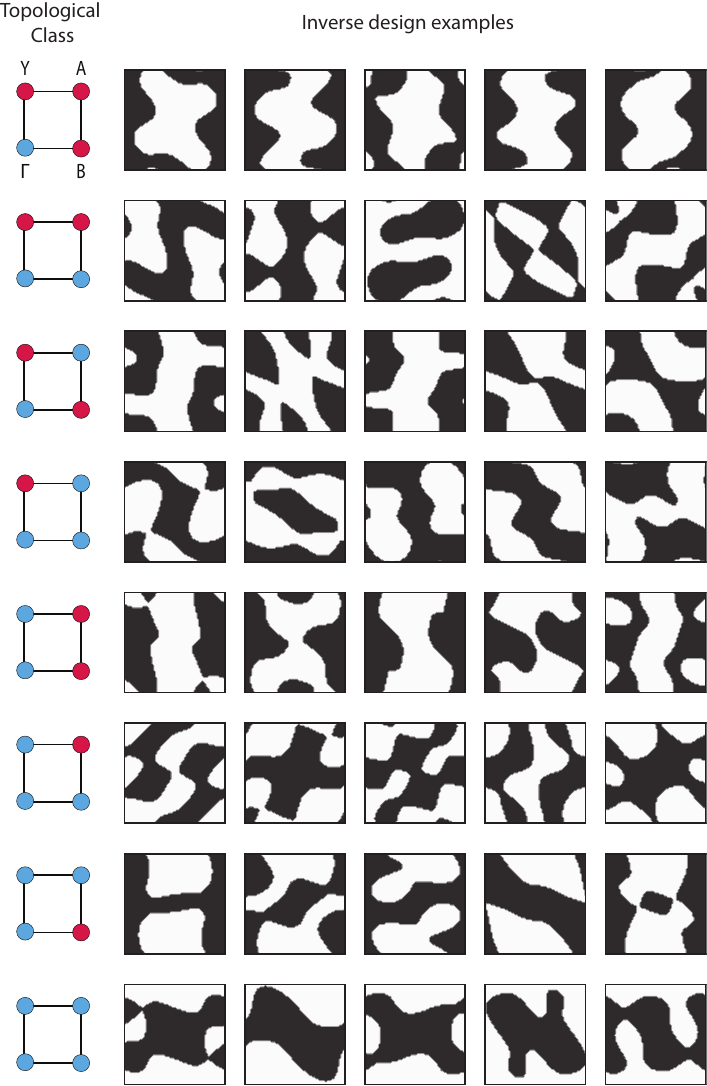}}
    \caption{\textbf{Example photonic crystals from inverse design:} To illustrate the diversity of our inverse designed dataset, we show here the real space dielectric functions corresponding to photonic crystals that were randomly sampled from this dataset. The lowest transverse magnetic band for each of these photonic crystals belongs to the corresponding topological class shown on the left. 
        }
    \label{figure: inverse design examples}
\end{figure}

To demonstrate that our inverse design method yields a diverse variety of photonic crystals, in \cref{figure: inverse design examples}, we show, for each topological category, five examples of photonic crystals unit cells that were created using our inverse design method. As is visually apparent, our inverse design method yields a significant diversity of dielectric profiles. This is because, prior to the optimization of our objective function (Equation. (4) of the main text), our Fourier coefficients are randomly initialized, ensuring that a large phase space of photonic crystals is explored. In addition, Fourier coefficients that are not important in determining the first band's topology (i.e. $\varepsilon_\mathbf{G}$ for $\mathbf{G}\neq \mathbf{G}_2, \mathbf{G}_1, \mathbf{G}_1+\mathbf{G}_2$) were randomly sampled and fixed at the outset of the inverse design process.


\end{document}

%% file: setup.tex
\usepackage[utf8]{inputenc}
\usepackage[english]{babel}

\usepackage{microtype} 
\usepackage{xspace} 

\usepackage{txfonts}  
\usepackage{txfontsb} 

\usepackage{bm} 

\usepackage{xcolor}
\usepackage[]{graphicx}

\usepackage[]{booktabs}
\usepackage{array}
\usepackage{layouts}
\usepackage{multirow}

\usepackage{enumerate}
\usepackage[inline]{enumitem}

\usepackage{xr}
\makeatletter
\newcommand*{\addFileDependency}[1]{
  \typeout{(#1)}
  \@addtofilelist{#1}
  \IfFileExists{#1}{}{\typeout{No file #1.}}
}
\makeatother

\usepackage{hyperref}
\hypersetup{colorlinks,
	linkcolor={blue!75!black!80!yellow},
	citecolor={blue!75!black!80!yellow},
	urlcolor={blue!75!black!80!yellow}
}

\usepackage[capitalize,nameinlink]{cleveref}

\crefname{subequations}{Eqs.}{Eqs.} 
\Crefname{subequations}{Eqs.}{Eqs.}
\crefformat{subequations}{#2Eqs.~(#1)#3}
\Crefformat{subequations}{#2Eqs.~(#1)#3}
\crefname{page}{p.}{p.} 

\usepackage{placeins}

\usepackage{siunitx}
\DeclareSIUnit[number-unit-product = ]\percent{\char`\%} 

\usepackage[centering,hmargin=16mm,tmargin=30mm,bmargin=26mm]{geometry}

\usepackage{textcomp} 
\usepackage{xifthen}
\usepackage{etoolbox}
\newboolean{togglecomments}
\newboolean{toggletodos}
\newboolean{togglechanges}

\setboolean{togglecomments}{true}
\setboolean{toggletodos}{true}
\setboolean{togglechanges}{false} 

\newcommand{\textblacksquare}{$\blacksquare$}
\newcommand{\todo}[1]{\ifbool{toggletodos}%
	{\textcolor{orange!80!yellow!95!black}{\small\textsf{{}\textsuperscript{\textsc{\textsf{todo}}}}[\ignorespaces#1]}} 
	{}}     
\newcommand{\comment}[2]{\ifbool{togglecomments}%
		{\textcolor{blue!70!black}{\small\sf\textsuperscript{\textsc{\textsf{#1}}}[\ignorespaces#2]}} 
		{}}     
\newcommand{\remove}[1]{\ifbool{togglechanges}
	{}    
	{\textcolor{red!70!black}{\ignorespaces#1}}}
\newcommand{\inset}[1]{\ifbool{togglechanges}
	{#1}  
	{\textcolor{green!70!black}{#1}}}
\newcommand{\citeremind}[1]{%
	[\textcolor{blue!75!black!80!yellow}{\textblacksquare%
		\ifthenelse{\isempty{#1}}{}{\textsuperscript{\tiny\textsf{\ignorespaces#1}}}%
	}]\xspace}

\input{commands.tex}

\makeatletter
\newcommand{\raisemath}[1]{\mathpalette{\raisem@th{#1}}}
\newcommand{\raisem@th}[3]{\raisebox{#1}{$#2#3$}}
\makeatother

\renewcommand{\paragraph}[1]{\vskip 1ex\noindent\textbf{#1.}~}

\usepackage[eulergreek]{sansmath}
\makeatletter
\renewcommand\@make@capt@title[2]{%
    \@ifx@empty\float@link{\@firstofone}{\expandafter\href\expandafter{\float@link}}%
    \sansmath\sffamily\textbf{#1\@caption@fignum@sep}#2 
}%

\makeatother


%% file: commands.tex
\newcommand{\br}[2]{(\mathrm{#1}\kern.125ex | \kern.125ex#2)}

\newcommand{\ie}{i.e.,\@\xspace} 

\newcommand{\appropto}{\mathrel{\vcenter{
			\offinterlineskip\halign{\hfil$##$\cr
				\propto\cr\noalign{\kern.2pt}\sim\cr\noalign{\kern-2.5pt}}}}}

\DeclareFontFamily{U}{mathx}{\hyphenchar\font45}
\DeclareFontShape{U}{mathx}{m}{n}{<5> <6> <7> <8> <9> <10>
                                  <10.95> <12> <14.4> <17.28> <20.74> <24.88>
                                  mathx10}{}
\DeclareSymbolFont{mathx}{U}{mathx}{m}{n}
\DeclareFontSubstitution{U}{mathx}{m}{n}

%% file: main.bib
@article{benalcazar2019quantization,
  title   = {Quantization of fractional corner charge in {$C_n$}-symmetric higher-order topological crystalline insulators},
  author  = {Benalcazar, Wladimir A and Li, Tianhe and Hughes, Taylor L},
  journal = {Phys. Rev. B},
  volume  = {99},
  number  = {24},
  pages   = {245151},
  year    = {2019},
  doi     = {10.1103/PhysRevB.99.245151}
}

@article{bradlyn2017topological,
  title     = {Topological quantum chemistry},
  author    = {Bradlyn, Barry and Elcoro, Luis and Cano, Jennifer and Vergniory, Maia G and Wang, Zhijun and Felser, Claudia and Aroyo, Mois I and Bernevig, B Andrei},
  journal   = {Nature},
  volume    = {547},
  number    = {7663},
  pages     = {298--305},
  year      = {2017},
  publisher = {Nature Publishing Group UK London},
  doi       = {10.1038/nature23268}
}

@article{christensen2020predictive,
  title     = {Predictive and generative machine learning models for photonic crystals},
  author    = {Christensen, Thomas and Loh, Charlotte and Picek, Stjepan and Jakobovi{\'c}, Domagoj and Jing, Li and Fisher, Sophie and Ceperic, Vladimir and Joannopoulos, John D and Solja{\v{c}}i{\'c}, Marin},
  journal   = {Nanophotonics},
  volume    = {9},
  number    = {13},
  pages     = {4183--4192},
  year      = {2020},
  publisher = {De Gruyter},
  doi       = {10.1515/nanoph-2020-0197}
}

@article{christensen2022location,
  title   = {Location and Topology of the Fundamental Gap in Photonic Crystals},
  author  = {Christensen, Thomas and Po, Hoi Chun and Joannopoulos, John D. and Solja{\v{c}}i{\'c}, Marin},
  journal = {Phys. Rev. X},
  volume  = {12},
  issue   = {2},
  pages   = {021066},
  year    = {2022},
  doi     = {10.1103/PhysRevX.12.021066}
}

@misc{Crystalline.jl,
  howpublished = {\url{https://github.com/thchr/Crystalline.jl}},
  title        = {{Crystalline.jl}~(v0.4.21)},
  urldate      = {2024-02-19},
  note         = {Accessed 19 February, 2024}
}

@article{haldane2008possible,
  title   = {Possible Realization of Directional Optical Waveguides in Photonic Crystals with Broken Time-Reversal Symmetry},
  author  = {Haldane, F. D. M. and Raghu, S.},
  journal = {Phys. Rev. Lett.},
  volume  = {100},
  issue   = {1},
  pages   = {013904},
  year    = {2008},
  doi     = {10.1103/PhysRevLett.100.013904}
}

@article{he2020quadrupole,
  title     = {Quadrupole topological photonic crystals},
  author    = {He, Li and Addison, Zachariah and Mele, Eugene J and Zhen, Bo},
  journal   = {Nature Commun.},
  volume    = {11},
  number    = {1},
  pages     = {3119},
  year      = {2020},
  publisher = {Nature Publishing Group UK London},
  doi       = {10.1038/s41467-020-16916-z}
}

@article{johnson2001block,
  title     = {Block-iterative frequency-domain methods for {Maxwell}'s equations in a planewave basis},
  author    = {Johnson, Steven G and Joannopoulos, John D},
  journal   = {Opt. Express},
  volume    = {8},
  number    = {3},
  pages     = {173--190},
  year      = {2001},
  publisher = {Optica Publishing Group},
  doi       = {10.1364/OE.8.000173}
}

@book{kaxiras2019quantum,
  title     = {Quantum theory of materials},
  author    = {Kaxiras, Efthimios and Joannopoulos, John D},
  year      = {2019},
  publisher = {Cambridge University Press}
}

@article{kim2023automated,
  doi     = {10.1021/acsphotonics.2c01866},
  year    = {2023},
  volume  = {10},
  pages   = {861},
  author  = {Samuel Kim and Thomas Christensen and Steven G. Johnson and Marin Solja{\v{c}}i{\'{c}}},
  title   = {Automated Discovery and Optimization of {3D} Topological Photonic Crystals},
  journal = {ACS Photonics}
}

@article{kruthoff2017topological,
  title    = {Topological Classification of Crystalline Insulators through Band Structure Combinatorics},
  author   = {Kruthoff, Jorrit and de Boer, Jan and van Wezel, Jasper and Kane, Charles L. and Slager, Robert-Jan},
  journal  = {Phys. Rev. X},
  volume   = {7},
  issue    = {4},
  pages    = {041069},
  numpages = {23},
  year     = {2017},
  doi      = {10.1103/PhysRevX.7.041069}
}

@article{lu2013weyl,
  title     = {{Weyl} points and line nodes in gyroid photonic crystals},
  author    = {Lu, Ling and Fu, Liang and Joannopoulos, John D and Solja{\v{c}}i{\'c}, Marin},
  journal   = {Nature Photonics},
  volume    = {7},
  number    = {4},
  pages     = {294--299},
  year      = {2013},
  publisher = {Nature Publishing Group UK London},
  doi       = {10.1038/nphoton.2013.42}
}

@article{lu2014topological,
  title   = {Topological photonics},
  author  = {Lu, L. and Joannopoulos, J. D. and Solja{\v{c}}i{\'c}, M.},
  journal = {Nat. Photon.},
  volume  = {8},
  number  = {11},
  pages   = {821-829},
  year    = {2014},
  doi     = {10.1038/nphoton.2014.248}
}

@misc{MPBUtils.jl,
  howpublished = {\url{https://github.com/thchr/MPBUtils.jl}},
  title        = {{MPBUtils.jl}~(v0.1.10)},
  urldate      = {2023-07-25},
  note         = {Accessed 25 July, 2023}
}

@article{ozawa2019topological,
  title     = {Topological photonics},
  author    = {Ozawa, Tomoki and Price, Hannah M. and Amo, Alberto and Goldman, Nathan and Hafezi, Mohammad and Lu, Ling and Rechtsman, Mikael C. and Schuster, David and Simon, Jonathan and Zilberberg, Oded and Carusotto, Iacopo},
  journal   = {Rev. Mod. Phys.},
  volume    = {91},
  issue     = {1},
  pages     = {015006},
  numpages  = {76},
  year      = {2019},
  month     = {Mar},
  publisher = {American Physical Society},
  doi       = {10.1103/RevModPhys.91.015006}
}

@article{po2017symmetry,
  title     = {Symmetry-based indicators of band topology in the 230 space groups},
  author    = {Po, Hoi Chun and Vishwanath, Ashvin and Watanabe, Haruki},
  journal   = {Nature Commun.},
  volume    = {8},
  number    = {1},
  pages     = {50},
  year      = {2017},
  publisher = {Nature Publishing Group UK London},
  doi       = {10.1038/s41467-017-00133-2}
}

@article{po2018fragile,
  title   = {Fragile topology and {Wannier} obstructions},
  author  = {Po, Hoi Chun and Watanabe, Haruki and Vishwanath, Ashvin},
  journal = {Phys. Rev. Lett.},
  volume  = {121},
  number  = {12},
  pages   = {126402},
  year    = {2018},
  doi     = {10.1103/PhysRevLett.121.126402}
}

@article{rechtsman2013photonic,
  title     = {Photonic {Floquet} topological insulators},
  author    = {Rechtsman, Mikael C and Zeuner, Julia M and Plotnik, Yonatan and Lumer, Yaakov and Podolsky, Daniel and Dreisow, Felix and Nolte, Stefan and Segev, Mordechai and Szameit, Alexander},
  journal   = {Nature},
  volume    = {496},
  number    = {7444},
  pages     = {196--200},
  year      = {2013},
  publisher = {Nature Publishing Group UK London},
  doi       = {10.1038/nature12066}
}

@article{skirlo2015experimental,
  title     = {Experimental observation of large {Chern} numbers in photonic crystals},
  author    = {Skirlo, Scott A and Lu, Ling and Igarashi, Yuichi and Yan, Qinghui and Joannopoulos, John and Solja{\v{c}}i{\'c}, Marin},
  journal   = {Phys. Rev. Lett.},
  volume    = {115},
  number    = {25},
  pages     = {253901},
  year      = {2015},
  publisher = {APS},
  doi       = {10.1103/PhysRevLett.115.253901}
}

@article{song2020fragile,
  title   = {Fragile phases as affine monoids: classification and material examples},
  author  = {Song, Zhi-Da and Elcoro, Luis and Xu, Yuan-Feng and Regnault, Nicolas and Bernevig, B. Andrei},
  journal = {Phys. Rev. X},
  volume  = {10},
  pages   = {031001},
  year    = {2020},
  doi     = {10.1103/PhysRevX.10.031001}
}

@article{vaidya2023response,
  author  = {Vaidya, Sachin and Rechtsman, Mikael C. and Benalcazar, Wladimir A.},
  title   = {Polarization and Weak Topology in {Chern} Insulators},
  journal = {Phys. Rev. Lett.},
  year    = {2024},
  volume  = {132},
  pages   = {116602},
  doi     = {10.1103/PhysRevLett.132.116602}
}

@article{vaidya2023topological,
  title   = {Topological Phases of Photonic Crystals under Crystalline Symmetries},
  author  = {Sachin Vaidya and Ali Ghorashi and Thomas Christensen and Mikael C. Rechtsman and Wladimir A. Benalcazar},
  year    = {2023},
  journal = {Phys. Rev. B},
  doi     = {10.1103/PhysRevB.108.085116},
  volume  = {108},
  pages   = {085116}
}

@article{wang2008reflection,
  title   = {Reflection-Free One-Way Edge Modes in a Gyromagnetic Photonic Crystal},
  author  = {Wang, Zheng and Chong, Y. D. and Joannopoulos, John D. and Solja\ifmmode \check{c}\else \v{c}\fi{}i\ifmmode \acute{c}\else \'{c}\fi{}, Marin},
  journal = {Phys. Rev. Lett.},
  volume  = {100},
  issue   = {1},
  pages   = {013905},
  year    = {2008},
  doi     = {10.1103/PhysRevLett.100.013905}
}

@article{wang2009observation,
  title     = {Observation of unidirectional backscattering-immune topological electromagnetic states},
  author    = {Wang, Zheng and Chong, Yidong and Joannopoulos, John D and Solja{\v{c}}i{\'c}, Marin},
  journal   = {Nature},
  volume    = {461},
  number    = {7265},
  pages     = {772--775},
  year      = {2009},
  publisher = {Nature Publishing Group UK London},
  doi       = {10.1038/nature08293}
}

@article{xie2018second,
  title     = {Second-order photonic topological insulator with corner states},
  author    = {Xie, Bi-Ye and Wang, Hong-Fei and Wang, Hai-Xiao and Zhu, Xue-Yi and Jiang, Jian-Hua and Lu, Ming-Hui and Chen, Yan-Feng},
  journal   = {Phys. Rev. B},
  volume    = {98},
  number    = {20},
  pages     = {205147},
  year      = {2018},
  publisher = {APS},
  doi       = {10.1103/PhysRevB.98.205147}
}

@article{jiang2021deep,
  title={Deep neural networks for the evaluation and design of photonic devices},
  author={Jiang, Jiaqi and Chen, Mingkun and Fan, Jonathan A},
  journal={Nature Reviews Materials},
  volume={6},
  number={8},
  pages={679--700},
  year={2021},
  publisher={Nature Publishing Group UK London},
  doi={10.1038/s41578-020-00260-1}
}

@article{liu2024kan,
  title={Kan: Kolmogorov-arnold networks},
  author={Liu, Ziming and Wang, Yixuan and Vaidya, Sachin and Ruehle, Fabian and Halverson, James and Solja{\v{c}}i{\'c}, Marin and Hou, Thomas Y and Tegmark, Max},
  journal={arXiv preprint arXiv:2404.19756},
  year={2024},
  doi={10.48550/arXiv.2404.19756
}
}

@article{ghorashi2024prevalence,
  title={Prevalence of two-dimensional photonic topology},
  author={Ghorashi, Ali and Vaidya, Sachin and Rechtsman, Mikael C and Benalcazar, Wladimir A and Solja{\v{c}}i{\'c}, Marin and Christensen, Thomas},
  journal={Physical review letters},
  volume={133},
  number={5},
  pages={056602},
  year={2024},
  publisher={APS},
  doi={10.1103/PhysRevLett.133.056602}
}

@article{zhang2018machine,
  title={Machine learning topological invariants with neural networks},
  author={Zhang, Pengfei and Shen, Huitao and Zhai, Hui},
  journal={Physical review letters},
  volume={120},
  number={6},
  pages={066401},
  year={2018},
  publisher={APS},
  doi={10.1103/PhysRevLett.120.066401}
}

@article{deng2024inverse,
  title={Inverse design in photonic crystals},
  author={Deng, Ruhuan and Liu, Wenzhe and Shi, Lei},
  journal={Nanophotonics},
  volume={13},
  number={8},
  pages={1219--1237},
  year={2024},
  publisher={De Gruyter},
  doi={10.1515/nanoph-2023-0750}
}

@article{pilozzi2018machine,
  title={Machine learning inverse problem for topological photonics},
  author={Pilozzi, Laura and Farrelly, Francis A and Marcucci, Giulia and Conti, Claudio},
  journal={Communications Physics},
  volume={1},
  number={1},
  pages={57},
  year={2018},
  publisher={Nature Publishing Group UK London},
  doi={10.1038/s42005-018-0058-8}
}

@article{chen2022inverse,
  title={Inverse design of photonic and phononic topological insulators: a review},
  author={Chen, Yafeng and Lan, Zhihao and Su, Zhongqing and Zhu, Jie},
  journal={Nanophotonics},
  volume={11},
  number={19},
  pages={4347--4362},
  year={2022},
  publisher={De Gruyter},
  doi={10.1515/nanoph-2022-0309}
}

@article{gupta2023tandem,
  title={Tandem neural network based design of multiband antennas},
  author={Gupta, Aggraj and Karahan, Emir Ali and Bhat, Chandan and Sengupta, Kaushik and Khankhoje, Uday K},
  journal={IEEE Transactions on Antennas and Propagation},
  volume={71},
  number={8},
  pages={6308--6317},
  year={2023},
  publisher={IEEE},
  doi={10.1109/TAP.2023.3276524}
}

@article{jiang2025machine,
  title={Machine Learning Inverse Design of Topological Quantum States in Photonic Topological Insulators},
  author={Jiang, Zhen and Wang, Yixin and Ji, Bo and He, Guangqiang and Jiang, Chun},
  journal={ACS Photonics},
  year={2025},
  publisher={ACS Publications},
  doi={10.1021/acsphotonics.4c02592}
}

@article{rawat2017deep,
  title={Deep convolutional neural networks for image classification: A comprehensive review},
  author={Rawat, Waseem and Wang, Zenghui},
  journal={Neural computation},
  volume={29},
  number={9},
  pages={2352--2449},
  year={2017},
  publisher={MIT Press},
  doi={10.1162/NECO_a_00990}
}

@inproceedings{guo2017simple,
  title={Simple convolutional neural network on image classification},
  author={Guo, Tianmei and Dong, Jiwen and Li, Henjian and Gao, Yunxing},
  booktitle={2017 IEEE 2nd International Conference on Big Data Analysis (ICBDA)},
  pages={721--724},
  year={2017},
  organization={IEEE},
  doi={10.1109/ICBDA.2017.8078730}
}

@book{kittel2018introduction,
  title={Introduction to solid state physics},
  author={Kittel, Charles and McEuen, Paul},
  year={2018},
  publisher={John Wiley \& Sons}
}

@article{siriwardane2022generative,
  title={Generative design of stable semiconductor materials using deep learning and density functional theory},
  author={Siriwardane, Edirisuriya M Dilanga and Zhao, Yong and Perera, Indika and Hu, Jianjun},
  journal={npj Computational Materials},
  volume={8},
  number={1},
  pages={164},
  year={2022},
  publisher={Nature Publishing Group UK London},
  doi={10.1038/s41524-022-00850-3}
}

@article{christiansen2021inverse,
  title={Inverse design in photonics by topology optimization: tutorial},
  author={Christiansen, Rasmus E and Sigmund, Ole},
  journal={Journal of the Optical Society of America B},
  volume={38},
  number={2},
  pages={496--509},
  year={2021},
  publisher={Optical Society of America},
  doi={10.1364/JOSAB.406048}
}

@article{aroyo2006bilbao,
  title={Bilbao Crystallographic Server: I. Databases and crystallographic computing programs},
  author={Aroyo, Mois Ilia and Perez-Mato, Juan Manuel and Capillas, Cesar and Kroumova, Eli and Ivantchev, Svetoslav and Madariaga, Gotzon and Kirov, Asen and Wondratschek, Hans},
  journal={Zeitschrift f{\"u}r Kristallographie-Crystalline Materials},
  volume={221},
  number={1},
  pages={15--27},
  year={2006},
  publisher={De Gruyter Oldenbourg},
  doi={10.1524/zkri.2006.221.1.15}
}

@article{ota2020active,
  title={Active topological photonics},
  author={Ota, Yasutomo and Takata, Kenta and Ozawa, Tomoki and Amo, Alberto and Jia, Zhetao and Kante, Boubacar and Notomi, Masaya and Arakawa, Yasuhiko and Iwamoto, Satoshi},
  journal={Nanophotonics},
  volume={9},
  number={3},
  pages={547--567},
  year={2020},
  doi={https://doi.org/10.1515/nanoph-2019-0376},
  publisher={De Gruyter}
}

@article{smirnova2020nonlinear,
  title={Nonlinear topological photonics},
  author={Smirnova, Daria and Leykam, Daniel and Chong, Yidong and Kivshar, Yuri},
  journal={Applied Physics Reviews},
  volume={7},
  number={2},
  year={2020},
  publisher={AIP Publishing},
  doi={https://doi.org/10.1063/1.5142397}
}

@article{parto2020non,
  title={Non-Hermitian and topological photonics: optics at an exceptional point},
  author={Parto, Midya and Liu, Yuzhou GN and Bahari, Babak and Khajavikhan, Mercedeh and Christodoulides, Demetrios N},
  journal={Nanophotonics},
  volume={10},
  number={1},
  pages={403--423},
  year={2020},
  publisher={De Gruyter},
  doi={https://doi.org/10.1515/nanoph-2020-0434}
}

@article{jalali2023topological,
  title={Topological photonics: Fundamental concepts, recent developments, and future directions},
  author={Jalali Mehrabad, Mahmoud and Mittal, Sunil and Hafezi, Mohammad},
  journal={Physical Review A},
  volume={108},
  number={4},
  pages={040101},
  year={2023},
  publisher={APS},
  doi={https://doi.org/10.1103/PhysRevA.108.040101}
}

@article{pontula2024non,
  title={Non-reciprocal frequency conversion in a multimode nonlinear system},
  author={Pontula, Sahil and Vaidya, Sachin and Roques-Carmes, Charles and Uddin, Shiekh Zia and Soljacic, Marin and Salamin, Yannick},
  journal={arXiv preprint arXiv:2409.14299},
  year={2024},
  doi={
https://doi.org/10.48550/arXiv.2409.14299}
}

@article{sloan2025noise,
  title={Noise immunity in quantum optical systems through non-Hermitian topology},
  author={Sloan, Jamison and Vaidya, Sachin and Rivera, Nicholas and Solja{\v{c}}i{\'c}, Marin},
  journal={arXiv preprint arXiv:2503.11620},
  year={2025},
  doi={
https://doi.org/10.48550/arXiv.2503.11620}
}

@article{schulz2021topological,
  title={Topological photonics in 3D micro-printed systems},
  author={Schulz, Julian and Vaidya, Sachin and J{\"o}rg, Christina},
  journal={APL Photonics},
  volume={6},
  number={8},
  year={2021},
  publisher={AIP Publishing},
  doi={https://doi.org/10.1063/5.0058478}
}

@article{kohn1965self,
  title={Self-consistent equations including exchange and correlation effects},
  author={Kohn, Walter and Sham, Lu Jeu},
  journal={Physical review},
  volume={140},
  number={4A},
  pages={A1133},
  year={1965},
  publisher={APS},
  doi={https://doi.org/10.1103/PhysRev.140.A1133}
}

@article{vaidya2020observation,
  title={Observation of a charge-2 photonic Weyl point in the infrared},
  author={Vaidya, Sachin and Noh, Jiho and Cerjan, Alexander and J{\"o}rg, Christina and Von Freymann, Georg and Rechtsman, Mikael C},
  journal={Physical review letters},
  volume={125},
  number={25},
  pages={253902},
  year={2020},
  publisher={APS},
  doi={https://doi.org/10.1103/PhysRevLett.125.253902}
}

@article{lyngby2022data,
  title={Data-driven discovery of 2D materials by deep generative models},
  author={Lyngby, Peder and Thygesen, Kristian Sommer},
  journal={npj Computational Materials},
  volume={8},
  number={1},
  pages={232},
  year={2022},
  publisher={Nature Publishing Group UK London},
  doi = {https://doi.org/10.1038/s41524-022-00923-3}
}

@article{davoodi2025active,
  title={Active physics-informed deep learning: surrogate modeling for nonplanar wavefront excitation of topological nanophotonic devices},
  author={Davoodi, Fatemeh},
  journal={Nano Letters},
  volume={25},
  number={2},
  pages={768--775},
  year={2025},
  publisher={ACS Publications},
  doi={https://doi.org/10.1021/acs.nanolett.4c05120}
}

@article{Slowlight_chern2,
  title={Broadband Topological Slow Light through Brillouin Zone Winding},
  author={Mann, Sander A and Al{\`u}, Andrea},
  journal={Physical Review Letters},
  volume={127},
  number={12},
  pages={123601},
  year={2021},
  publisher={APS},
  doi={https://doi.org/10.1103/PhysRevLett.127.123601}
}

@article{Slowlight_chern1,
  title={Broadband topological slow light through higher momentum-space winding},
  author={Guglielmon, Jonathan and Rechtsman, Mikael C},
  journal={Physical Review Letters},
  volume={122},
  number={15},
  pages={153904},
  year={2019},
  publisher={APS},
  doi={https://doi.org/10.1103/PhysRevLett.122.153904}
}

@article{Slowlight_chern3,
  title={Topological slow light via coupling chiral edge modes with flatbands},
  author={Yu, Letian and Xue, Haoran and Zhang, Baile},
  journal={Applied Physics Letters},
  volume={118},
  number={7},
  pages={071102},
  year={2021},
  publisher={AIP Publishing LLC},
  doi={https://doi.org/10.1063/5.0039839}
}

@article{wang2025topological,
  title={Topological microwave isolator with $>$ 100-dB isolation},
  author={Wang, Gang and Lu, Ling},
  journal={Nature Photonics},
  pages={1--6},
  year={2025},
  publisher={Nature Publishing Group UK London},
  doi={https://doi.org/10.1038/s41566-025-01750-w}
}

@article{dirac_laser1,
  title={Larger-area single-mode photonic crystal surface-emitting lasers enabled by an accidental {Dirac} point},
  author={Chua, Song-Liang and Lu, Ling and Bravo-Abad, Jorge and Joannopoulos, John D and Solja{\v{c}}i{\'c}, Marin},
  journal={Optics Letters},
  volume={39},
  number={7},
  pages={2072--2075},
  year={2014},
  publisher={Optical Society of America},
  doi={https://doi.org/10.1364/OL.39.002072}
}

@article{dirac_laser2,
  title={Enabling single-mode behavior over large areas with photonic {Dirac} cones},
  author={Bravo-Abad, Jorge and Joannopoulos, John D and Solja{\v{c}}i{\'c}, Marin},
  journal={Proceedings of the National Academy of Sciences},
  volume={109},
  number={25},
  pages={9761--9765},
  year={2012},
  publisher={National Acad Sciences},
  doi={https://doi.org/10.1073/pnas.1207335109}
}

@article{BerkSEL,
  title={Scalable single-mode surface-emitting laser via open-Dirac singularities},
  author={Contractor, Rushin and Noh, Wanwoo and Redjem, Walid and Qarony, Wayesh and Martin, Emma and Dhuey, Scott and Schwartzberg, Adam and Kant{\'e}, Boubacar},
  journal={Nature},
  volume={608},
  number={7924},
  pages={692--698},
  year={2022},
  publisher={Nature Publishing Group UK London},
  doi={https://doi.org/10.1038/s41586-022-05021-4}
}

@article{diracexp1,
author = {Lei Hu and Kang Xie and Zhijia Hu and Qiuping Mao and Jiangying Xia and Haiming Jiang and Junxi Zhang and Jianxiang Wen and Jingjing Chen},
journal = {Opt. Express},
number = {7},
pages = {8213--8223},
publisher = {OSA},
title = {Experimental observation of wave localization at the {Dirac} frequency in a two-dimensional photonic crystal microcavity},
volume = {26},
month = {Apr},
year = {2018},
url = {http://www.opticsexpress.org/abstract.cfm?URI=oe-26-7-8213},
doi = {10.1364/OE.26.008213},
}

@article{diracexp2,
title = {Observation of {Dirac} mode in modified honeycomb hollow core photonic crystal fiber},
journal = {Optical Materials},
volume = "89",
pages = "203--208",
year = "2019",
issn = "0925-3467",
doi = "https://doi.org/10.1016/j.optmat.2019.01.029",
url = "http://www.sciencedirect.com/science/article/pii/S0925346719300527",
author = "Li Yang and Guoquan Qian and Guowu Tang and Fangqiang Yuan and Zhishen Zhang and Kaimin Huang and Zhenguo Shi and Qi Qian and Zhongmin Yang"
}

@article{diracmode1,
author = {Xie, Kang and Jiang, Haiming and Boardman, Allan D. and Liu, Yong and Wu, Zhenhai and Xie, Ming and Jiang, Ping and Xu, Quan and Yu, Ming and Davis, Lionel E.},
title = {Trapped photons at a {Dirac} point: a new horizon for photonic crystals},
journal = {Laser \& Photonics Reviews},
volume = {8},
number = {4},
pages = {583-589},
doi = {https://doi.org/10.1002/lpor.201300186},
year = {2014}
}

@article{diracexp3,
author={Xie, Kang
and Zhang, Wei
and Boardman, Allan D.
and Jiang, Haiming
and Hu, Zhijia
and Liu, Yong
and Xie, Ming
and Mao, Qiuping
and Hu, Lei
and Li, Qian
and Yang, Tianyu
and Wen, Fei
and Wang, Erlei},
title={Fiber guiding at the {Dirac} frequency beyond photonic bandgaps},
journal={Light: Science {\&} Applications},
year={2015},
month={Jun},
day={01},
volume={4},
number={6},
pages={e304-e304},
issn={2047-7538},
doi={10.1038/lsa.2015.77},
url={https://doi.org/10.1038/lsa.2015.77}
}

@article{vaidya2021point,
  title={Point-defect-localized bound states in the continuum in photonic crystals and structured fibers},
  author={Vaidya, Sachin and Benalcazar, Wladimir A and Cerjan, Alexander and Rechtsman, Mikael C},
  journal={Physical Review Letters},
  volume={127},
  number={2},
  pages={023605},
  year={2021},
  publisher={APS},
  doi={https://doi.org/10.1103/PhysRevLett.127.023605}
}

@article{jurgensen2021quantized,
  title={Quantized nonlinear Thouless pumping},
  author={J{\"u}rgensen, Marius and Mukherjee, Sebabrata and Rechtsman, Mikael C},
  journal={Nature},
  volume={596},
  number={7870},
  pages={63--67},
  year={2021},
  publisher={Nature Publishing Group UK London}, 
  doi={https://doi.org/10.1038/s41586-021-03688-9}
}

@article{mukherjee2021observation,
  title={Observation of unidirectional solitonlike edge states in nonlinear Floquet topological insulators},
  author={Mukherjee, Sebabrata and Rechtsman, Mikael C},
  journal={Physical Review X},
  volume={11},
  number={4},
  pages={041057},
  year={2021},
  publisher={APS},
  doi={https://doi.org/10.1103/PhysRevX.11.041057}
}

@article{joannopoulos2008molding,
  title={Molding the flow of light},
  author={Joannopoulos, John D and Johnson, Steven G and Winn, Joshua N and Meade, Robert D},
  journal={Princet. Univ. Press. Princeton, NJ [ua]},
  volume={12},
  pages={33},
  year={2008}
}

@misc{Sparse-Networks-For-Topological-Photonics,
  howpublished = {\url{https://github.com/AliGhorashiCMT/Sparse-networks-for-topological-photonics}},
  urldate      = {2025-11-23},
  note       = {Accessed 23 November, 2025}
}
